\def\hii{H~{\sc ii}}

\documentclass[useAMS,usenatbib,epsfig]{mn2e}

\usepackage{rotating}
\usepackage{float,amssymb}
\usepackage{natbib}
\usepackage{verbatim}
\usepackage{hyperref,graphicx}

\title[X-shooter GRB high-$z$ extinction curves]{X-shooting GRBs at high redshift: probing dust production history\thanks{Based on observations collected at the European Organisation for Astronomical Research in the Southern Hemisphere, 8.2 m Very Large Telescope (VLT) with the X-shooter instrument mounted at UT2 under ESO programmes 087.A-0055(B), 088.A-0051(B), 089.A-0067(B), 090.A-0088(C), 092.D-0633(E), 093.A-0069(A), 096.A-0079(A), and 098.A-0055(A).}}
\author[T. Zafar et al.] {T. Zafar$^1$\thanks{e-mail: tayyaba.zafar@aao.gov.au}, P. M{\o}ller$^2$, D. Watson$^3$, J. Lattanzio$^4$, A. M. Hopkins$^1$, A. Karakas$^4$,
\newauthor
J. P. U. Fynbo$^3$,  N. R. Tanvir$^5$, J. Selsing$^3$, P. Jakobsson$^6$, K. E. Heintz$^{5, 3}$,
\newauthor
D. A. Kann$^{7}$, B. Groves$^{8}$, V. Kulkarni$^9$, S. Covino$^{10}$, V. D'Elia$^{11}$, J. Japelj$^{12}$,
\newauthor
D. Corre$^{13}$, S. Vergani$^{14}$ \\
$^1$ Australian Astronomical Observatory, PO Box 915, North Ryde, NSW 1670, Australia. \\
$^2$ European Southern Observatory, Karl-Schwarzschild-Strasse 2, 85748, Garching, Germany. \\
$^3$ The Cosmic Dawn Center, Niels Bohr Institute, University of Copenhagen, Juliane Maries Vej 30, DK-2100 Copenhagen, Denmark. \\
$4$ Monash centre for Astrophysics, School of Physics and Astronomy,10 College Walk, Monash University 3800, Australia \\
$^5$ Department of Physics and Astronomy, University of Leicester, University Road, Leicester, LE1 7RH, UK \\
$^6$ Centre for Astrophysics and Cosmology, Science Institute, University of Iceland, Dunhagi 5, 107 Reykjav\'ik, Iceland.\\
$^{7}$ Instituto de Astrof\'isica de Andaluc\'ia (IAA-CSIC), Glorieta de la Astronom\'ia s/n, E-18008, Granada, Spain. \\
$^8$ Research School of Astronomy and Astrophysics, Australian National University, Canberra, ACT 2611, Australia. \\
$^9$ University of South Carolina, Dept. of Physics and Astronomy, Columbia, SC 29208 \\
$^{10}$ Osservatorio Astronomico di Brera, via Bianchi 46, 23807, Merate (LC), Italy.\\
$^{11}$ Space Science Data Center - Agenzia Spaziale Italiana , Via del Politecnico snc., I-00133 Roma, Italy.\\
$^{12}$ Astronomical Institute Anton Pannekoek, University of Amsterdam, Science Park 904, 1098 XH, Amsterdam, the Netherlands. \\
$^{13}$ Aix Marseille Universit\'e, CNRS, LAM, Laboratoire d'Astrophysique de Marseille, Marseille, France \\
$^{14}$ GEPI, Observatoire de Paris, PSL Universit\'e, CNRS, 5 Place Jules Janssen, 92190 Meudon,
France.
}

\begin{document}


\pagerange{\pageref{firstpage}--\pageref{lastpage}} \pubyear{2018}

\maketitle

\label{firstpage}

\begin{abstract}
Evolved asymptotic giant branch (AGB) stars and Type\,Ia supernovae (SNe) are important contributors to the elements that form dust in the interstellar medium of galaxies, in particular, carbon and iron. However, they require at least a Gyr to start producing these elements, therefore, a change in dust quantity or properties may appear at high redshifts. In this work, we use extinction of $\gamma$-ray burst (GRB) afterglows as a tool to look for variations in dust properties at $z\ge3$. We use a spectroscopically selected sample of GRB afterglows observed with the VLT/X-shooter instrument to determine extinction curves out to high redshifts. We present ten new $z\ge3$ X-shooter GRBs of which six are dusty. Combining these with individual extinction curves of three previously known $z\ge3$ GRBs, we find an average extinction curve consistent with the SMC-Bar. A comparison with spectroscopically selected GRBs at all redshifts indicates a drop in visual extinction ($A_V$) at $z>3.5$ with no moderate or high extinction bursts. We check for observational bias using template spectra and find that GRBs up to $z\sim8$ are detectable with X-shooter up to $A_V\sim0.3$\,mag. Although other biases are noted, a uniformly low dust content above $z>3.5$ indicates a real drop, suggesting a transition in dust properties and/or available dust building blocks. The remarkable increase in dust content at $z<3.5$ could arise due to carbon and possibly iron production by the {\it first} carbon-rich AGB and Type\,Ia SNe, respectively. Alternatively, $z>3.5$ dust drop could be the result of low stellar masses of GRB host galaxies.
\end{abstract}

\section{Introduction}
The evolution of dust in the early Universe is highly debated. Different formation mechanisms have been proposed to dominate the dust production at high redshifts, including core-collapse supernovae \citep[CCSNe;][]{todini01,morgan03,hirashita05,dwek07}, massive asymptotic giant branch (AGB) stars \citep{valiante09,hirashita14}, and interstellar medium (ISM) grain growth \citep{draine09,michalowski15,mattsson14}. All of these mechanisms suffer from serious difficulties: $i)$ Formation and dust production of AGB stars takes too long to explain $z>7$ dusty galaxies \citep{michalowski15}, and also they appear to be minor dust
contributors even in the local universe \citep{matsuura09,meixner06}, $ii)$ ISM grain growth needs to be extremely rapid, and may be hindered by the formation of surface ices \citep{ferrara17}, and $iii)$ CCSNe require time to build up dust and metals. While CCSNe are clearly shown to produce dust \citep{matsuura09,gomez12,delooze16,wesson10,bevan16,temim17}, CCSN dust production is at the same time expected to be counteracted by the destructive effects of supernova shocks \citep{nozawa07,schneider12}, apparently destroying dust so efficiently that the existence of dusty, star-forming galaxies in the early universe \citep[e.g.][]{cooray14,watson15,laporte17} is something of a mystery \citep{michalowski12,rowlands14}. While there are potentially ways out of some of these difficulties \citep[e.g.][]{gall14}, it is obviously vital to constrain observationally the nature of the dust at all redshifts to solve this mystery.

The energetic afterglows of long-duration $\gamma$-ray bursts (GRBs) are powerful probes with which to study the ISM of galaxies into the epoch of reionisation \citep[e.g.,][]{tanvir09,gehrels09}. Their association with the explosions of massive stars \citep[e.g.,][]{woosley93,galama98,hjorth03,cano17} and their simple spectral shapes \citep{sari98,granot02} make them unique probes to study dust extinction in star-forming regions at cosmological distances \citep[e.g.,][]{watson06,kann10,greiner11,zafar11,schady12,zafar18}. While quasars can also be used to determine extinction, their selection criteria strongly favour low extinction sightlines, and the complexity of quasar spectra and the uncertainty of their intrinsic slopes leads to a degeneracy in derived dust properties, especially at high redshifts \citep[e.g.][]{maiolino04,gallerani10,hjorth13}.

Currently, average extinction laws from the Milky Way (MW) and the Large and Small Magellanic Clouds (LMC and SMC) are often used to quantify dust properties of the environments that existed around or at the epoch of reionisation \citep{zafar12,tanvir17,zafar18b}. It is important to derive the individual extinction curves of high redshift sources to determine the grain properties and content of dust in the early Universe and to understand the appearance and formation of the first stars and galaxies. It is important here to recall that in case of a simple back-lit dust screen, an extinction curve infers the light loss due to scattering and absorption by the dust. In the more commonly encountered situation of an integrated geometry where the dust is mixed with the stars and ionised gas, an attenuation curve defines the relative distribution of dust and emitting sources. For star-forming galaxies, the effective reddening of the galaxy is determined by a Calzetti attenuation law \citep{calzetti94} which is {\it greyer} than the MW, LMC, and SMC extinction curves. However, attenuation law through dust mixed with emitting sources and with SMC-type dust can lead to Calzetti-law attenuation \citep{gordon97,inoue05}.

The Neil Gehrels \emph{Swift} Observatory \citep{gehrels04}, enables discovery of GRBs up to the epoch of reionisation \citep{tanvir09,tanvir17}. The fast response and tremendous efforts of the follow-up teams ensure that simultaneous ground-based photometric and spectroscopic data are acquired to affirm high-redshift GRB afterglow discoveries. In the present generation of instruments, the VLT/X-shooter spectrograph \citep{vernet11} with its sensitivity, medium resolution, and wide-band from the ultraviolet (UV) to the near-infrared (NIR) is highly efficient and suitable to study dust properties of high redshift GRBs. Modelling of the X-ray to the NIR spectroscopic spectral energy distributions (SEDs) provides individual extinction curves of GRB afterglows to understand dust properties at high redshifts.

Previously, \citet{zafar11} reported a sudden drop in dust content in GRB afterglows above $z\ge4$ with a lack of $A_V\sim0.3$\,mag extinguished events. Recently, \citet{bolmer17} found that on average GRBs at $z>4$ contain less dust than at $z\sim2$. They claim it is an observational bias to not easily detect bursts with $A_V>0.5$\,mag at $z>4$ with the Gamma-Ray burst Optical and Near-Infrared Detector (GROND) mounted at the 2.2-m MPG telescope. Such a drop in dustiness at $z\sim4$ is also reported by \citet{mclure13,dunlop17} through the Atacama Large Millimeter Array (ALMA) observations of the Hubble Ultra Deep Field and through SCUBA-2 observations on the James Clerk Maxwell Telescope \citep{koprowski17}. A similar drop in the fraction of dusty ultra-luminous infrared galaxies (ULIRGs) in Ly$\alpha$ emission selected samples at redshifts larger than 2.5 was reported by \citet{nilsson09a}, while at the same time a $z=7.5$ faint dusty galaxy with $A_{\rm 1600}=1$\,mag has been confirmed \citep{watson15}. \citet{atek14} and \citet{hayes11} reported a decrease in dust and increase in Ly$\alpha$ escape fraction with increasing redshift up to $z\sim6$ for Ly$\alpha$ emitters, a decrease which is detectable already between redshifts of 2 and 3 \citep{nilsson09b}. There is thus a large body of evidence for an evolution of the dust content of the Universe with cosmic time. Here we aim to investigate the exact redshift of the drop in dust content and how large the variation is. This is done by measuring the amount of dust in GRBs at $z\ge3$ to pinpoint the epoch of the transition phase.

In this work, we selected GRBs above $z\geq3$ that were observed with the X-shooter instrument and have simultaneous photometric data available. In \S2 we present our X-shooter high redshift sample and provide details about the multi-wavelength data taken for each case. In \S3 we describe our dust model and SED analysis. The results and discussions are provided in \S4 and \S5. The conclusions of our high redshift dust analysis are given in \S6. Throughout the paper, errors denote 1$\sigma$ uncertainties and 3$\sigma$ limits are provided.

\begin{table}
\begin{minipage}[t]{\columnwidth}
\caption{The X-shooter $z\geq3$ GRB afterglow sample. The columns provide the GRB name, redshift, Galactic extinction, total Galactic equivalent neutral hydrogen column density, and mid-time of the afterglow SED (photometric mid-points).}
\label{grb:list}
\centering
\renewcommand{\footnoterule}{}  
\setlength{\tabcolsep}{8pt}
\begin{tabular}{l c c c c }\hline\hline
GRB & $z$ & $E(B-V)_{\rm Gal}$ & $N_{\rm H, Gal}$ & $\Delta t$ \\
	&  & mag &  10$^{20}$ cm$^{-2}$ &  hours \\
\hline\hline
110818A & 	3.360 & 0.03 & 2.93 & 6.200 \\
111123A & 	3.152 & 0.05 & 6.90 & 13.37 \\
120712A & 	4.175 & 0.04 & 4.12 & 10.43 \\
121201A & 	3.385 & 0.01 & 2.05 & 12.00 \\
130408A & 	3.758 & 0.22 & 32.0 & 1.500 \\
140311A & 	4.954 & 0.03 & 2.80 & 27.89 \\
140515A & 	6.327 & 0.02 & 2.54 & 16.83 \\
140614A & 	4.233 & 0.11 & 12.6 & 3.150 \\
151027B & 	4.062 & 0.18 & 9.43 & 8.011 \\
170202A & 	3.645 & 0.02 & 1.98 & 16.00 \\
\hline
\end{tabular}
\end{minipage}
\end{table}

\section{Ensemble selection}
A large sample of GRB afterglow spectra has been acquired with the VLT/X-shooter instrument under target of opportunity (ToO) programs. From March 2009 until March 2017, 121 spectra have been taken with X-shooter, eight of these being short-duration GRBs and the remaining 113 are long duration bursts. X-shooter has three spectroscopic arms (UVB: 300-550\,nm, VIS: 550-1,000\,nm, and NIR: 1,000-2,500\,nm) and for these arms, GRB afterglow spectra are usually taken with 1.0$''$ (UVB), 0.9$''$ (VIS), and 0.9$''$ (NIR) slit widths. The afterglow spectra are reduced and flux calibrated using the standard X-shooter pipeline (version 2.0; \citealt{modigliani10}). More details on the reduction and flux calibration, including background subtraction and extraction of each GRB afterglow observed under the X-shooter GRB legacy sample are provided in \citet{selsing18}.

Out of 113 long-duration GRBs, only 20 GRBs were discovered at $z\geq3$ \citep{selsing18}. The SED analysis of six $z\geq3$ GRB afterglows has already been presented in \citet{zafar18}. In addition, one case was excluded as being a host galaxy observation. We targeted the remaining 13 GRB afterglows to perform SED analysis which relies on robust flux calibration. Due to the usage of a broader 5.0$''$ wide slit for the flux standard star observations and malfunction of the Atmospheric Dispersion Correctors (ADC) for the UVB and VIS arms, the GRB afterglow spectra have sub-optimal flux calibration primarily due to slit-losses. However, note that after the ADC malfunctioning, the observations were always taken at the parallactic angle to minimise any effect of differential slit loss \citep[see][for more details]{selsing18}. We required photometric data around the X-shooter observations to have optimal flux calibration for each case \citep[see][]{japelj15}. Our dedicated search in the literature resulted in finding 10 new $z\geq3$ GRBs with simultaneous photometric observations. We constructed the SEDs for each of these 10 cases to derive individual extinction curves at $z\ge3$.

\begin{figure*}
  \centering
{\includegraphics[width=\columnwidth,clip=]{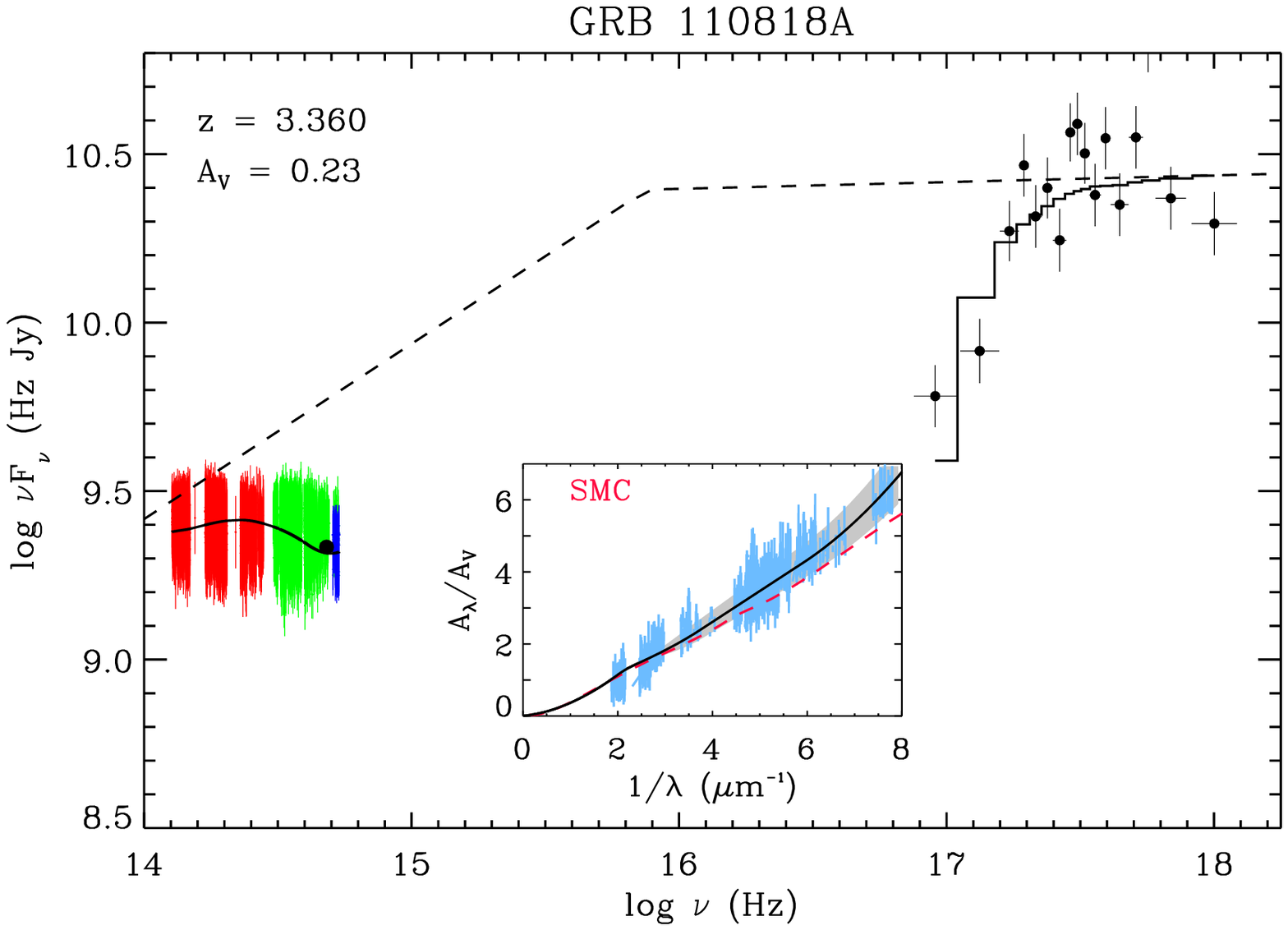}}
{\includegraphics[width=\columnwidth,clip=]{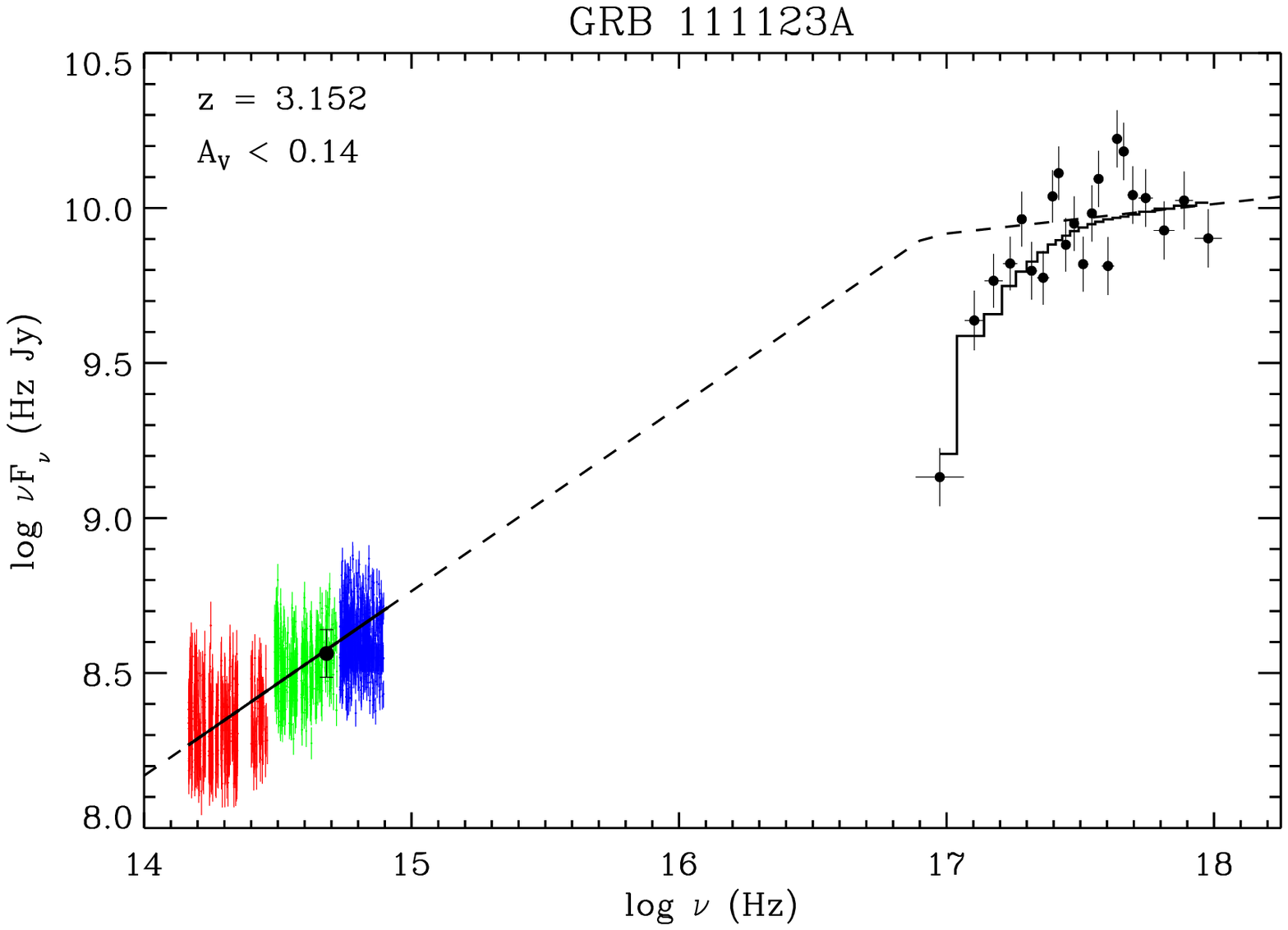}}
{\includegraphics[width=\columnwidth,clip=]{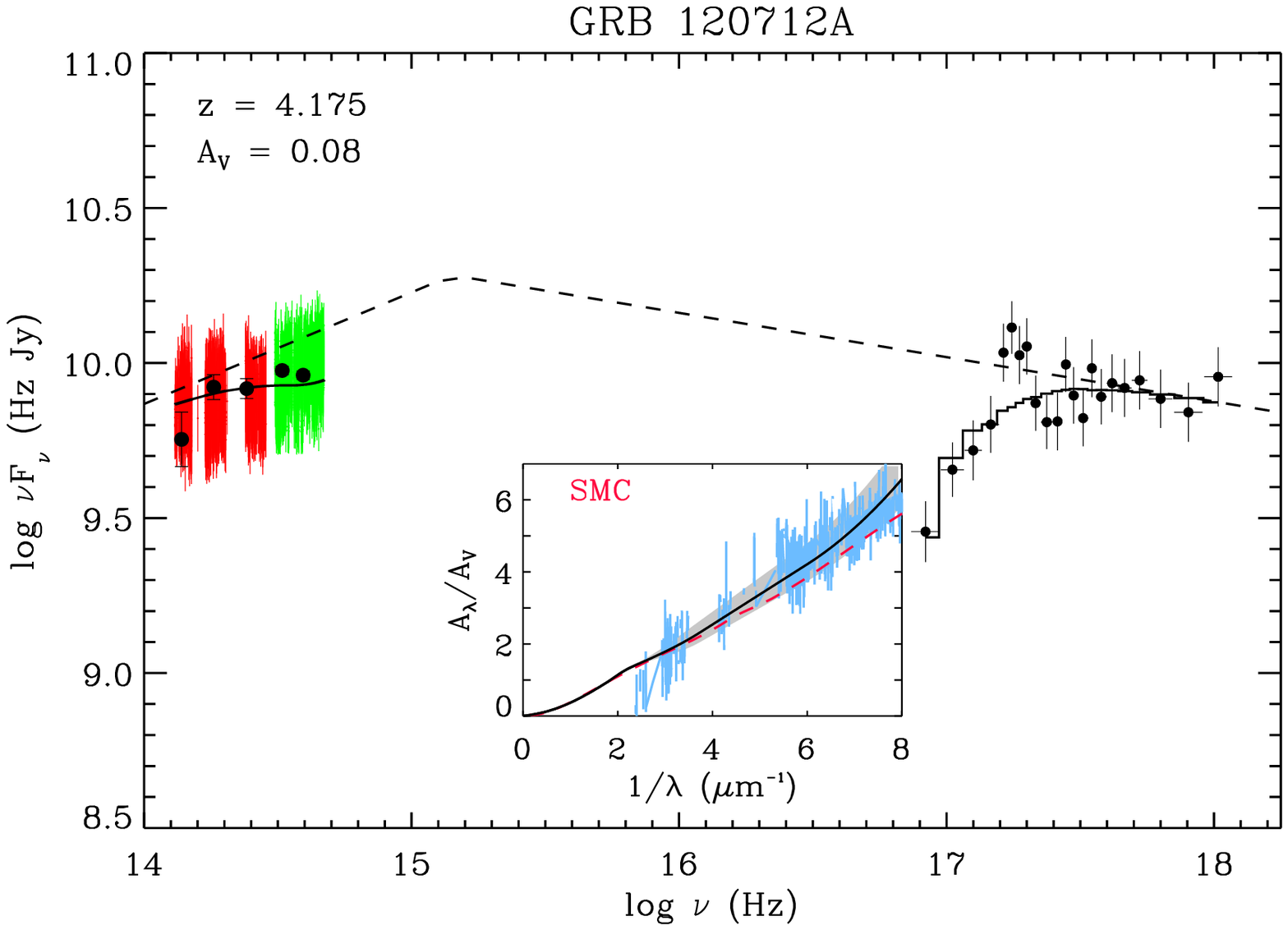}}
{\includegraphics[width=\columnwidth,clip=]{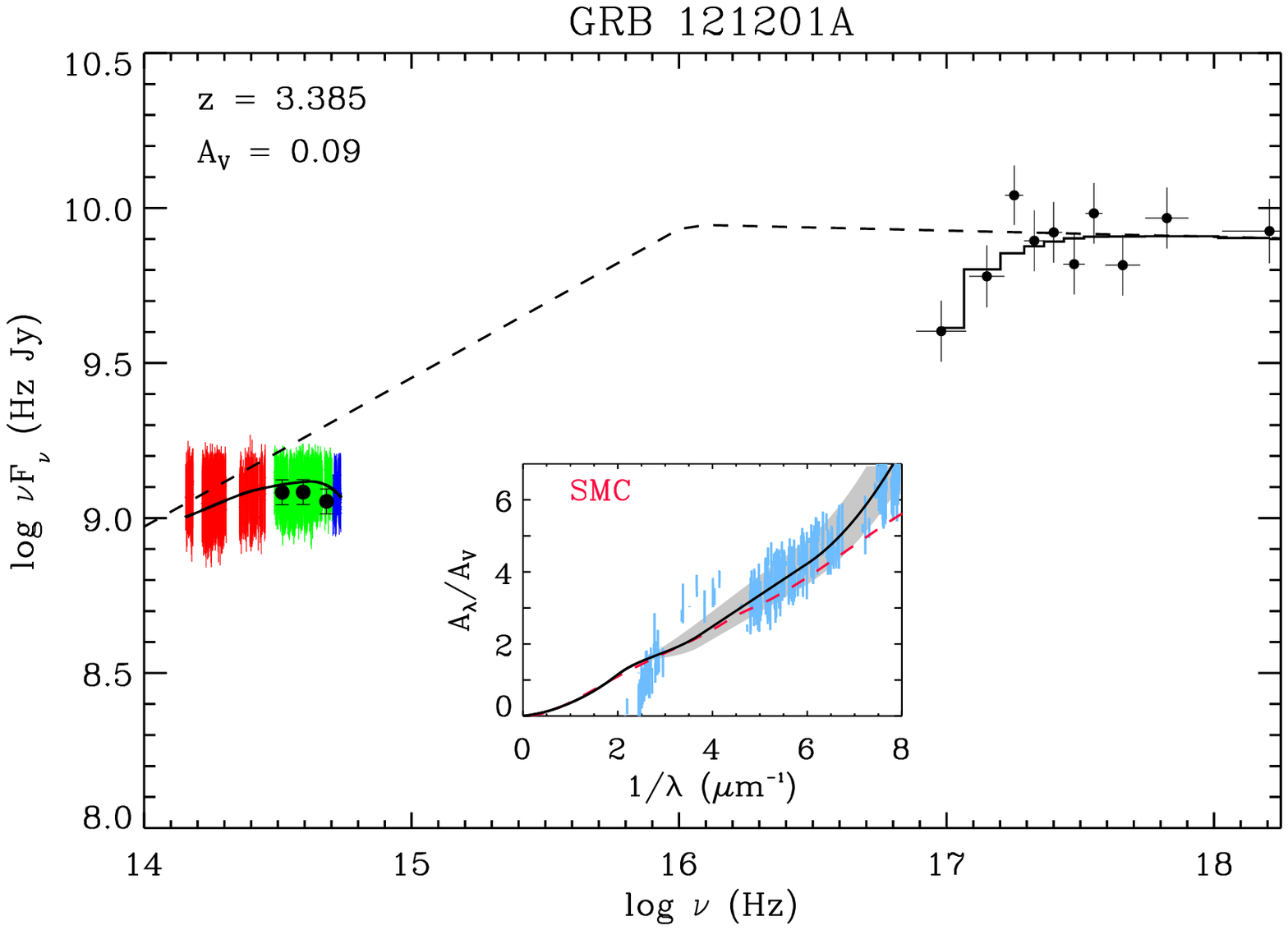}}
{\includegraphics[width=\columnwidth,clip=]{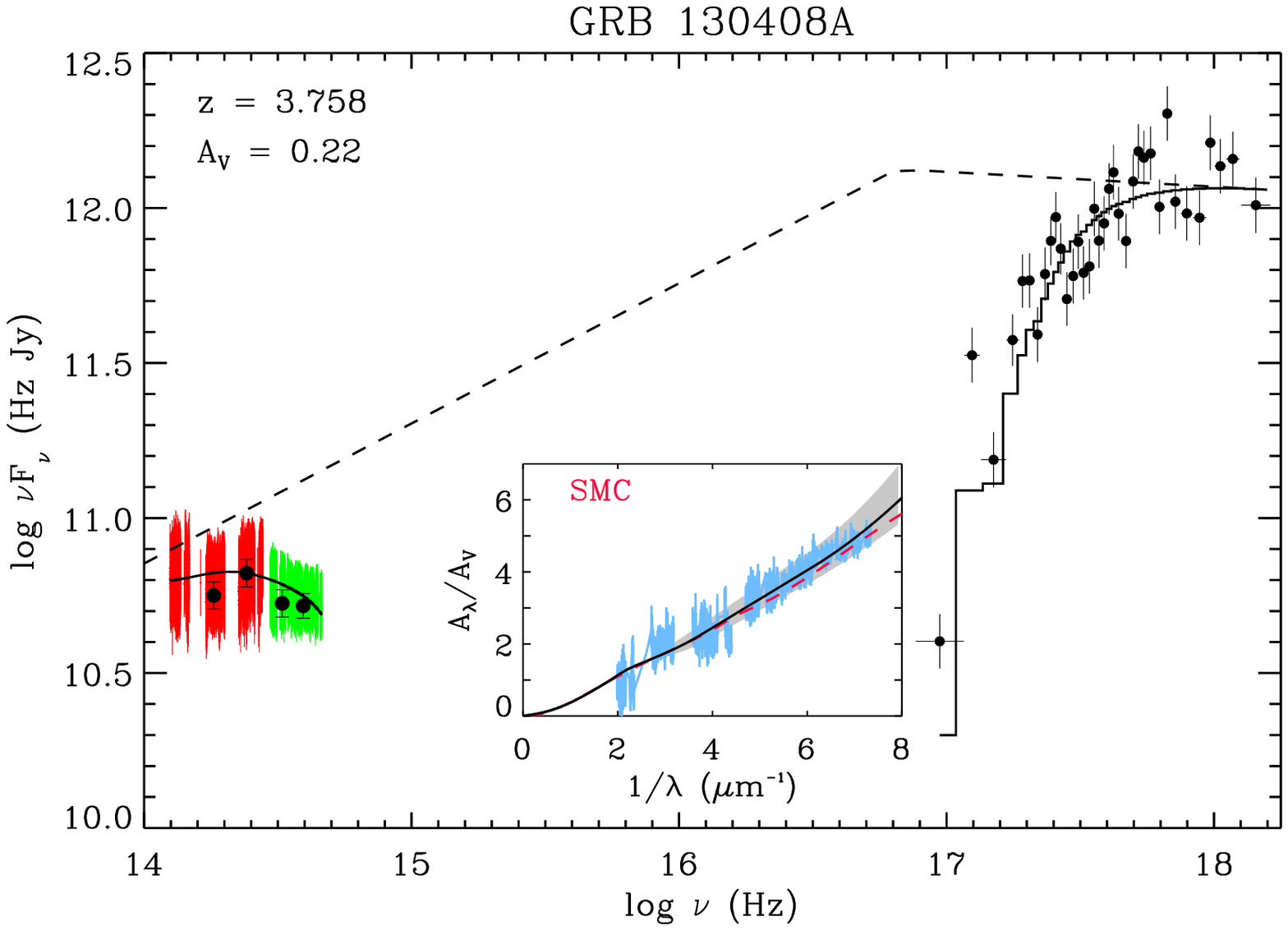}}
 {\includegraphics[width=\columnwidth,clip=]{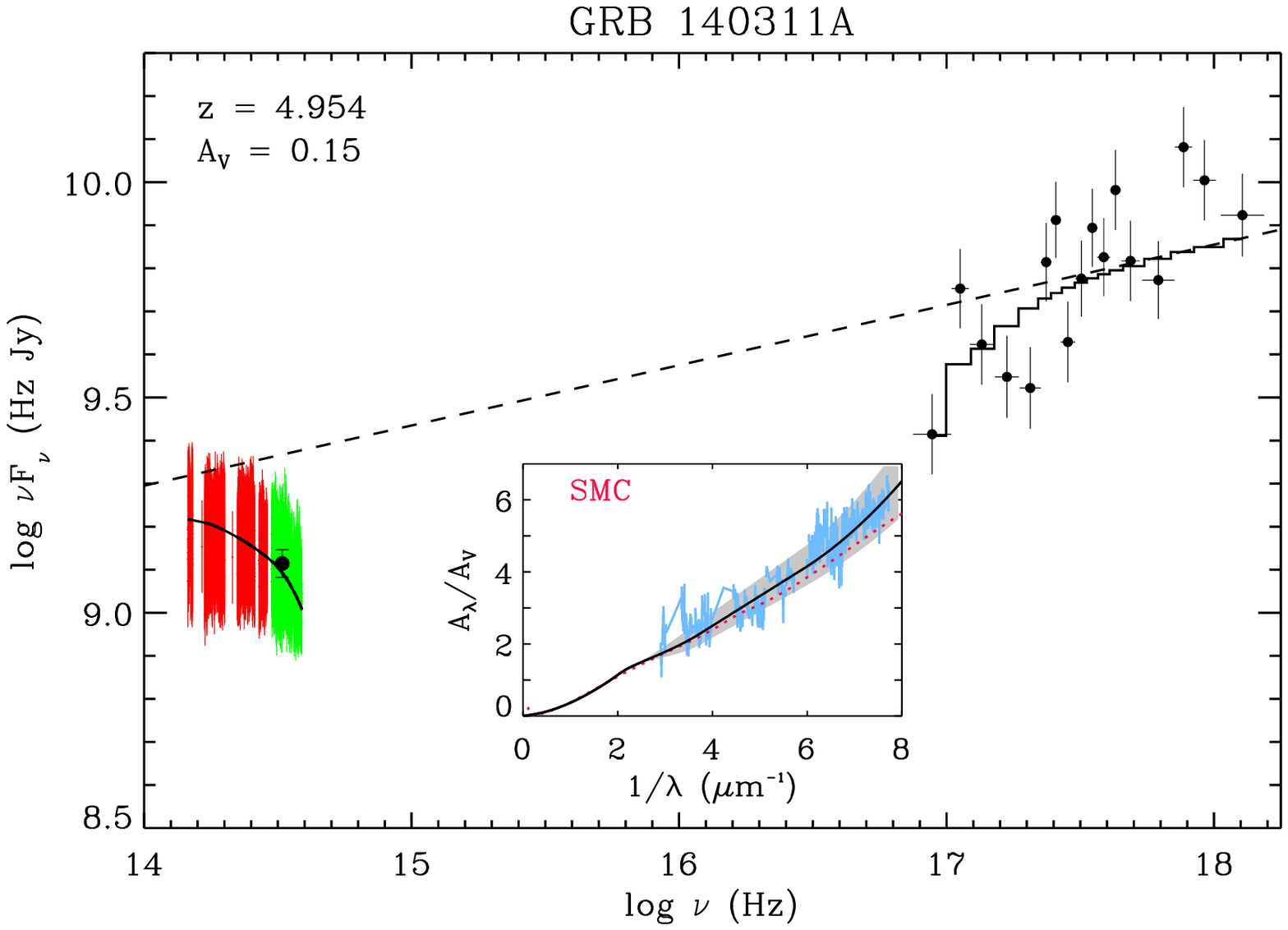}}
    \caption{The observer-frame X-shooter $z\geq3$ GRB afterglow SEDs and their best fit models and extinction curves. The \emph{Swift} X-ray data on the right is shown by black points. Towards the left side, X-shooter UVB (blue), VIS (green), and NIR (red) spectra and multi-band photometry (black points) together with errors are presented. The best fit dust and soft X-ray absorbed (solid lines) and extinction and X-ray absorption corrected spectral models (dashed lines) are shown in black. {\it Inset:} Only for dusty cases, the absolute extinction curves of the GRB afterglows are shown in black lines with grey-shading corresponding to the 1$\sigma$ uncertainty of the curves. The X-shooter spectra are represented by cyan curves. The typical SMC law from \citet{pei92} is shown as a red dashed line.}
         \label{grb:fits}
  \end{figure*}

\addtocounter{figure}{-1}
\begin{figure*}
  \centering
{\includegraphics[width=\columnwidth,clip=]{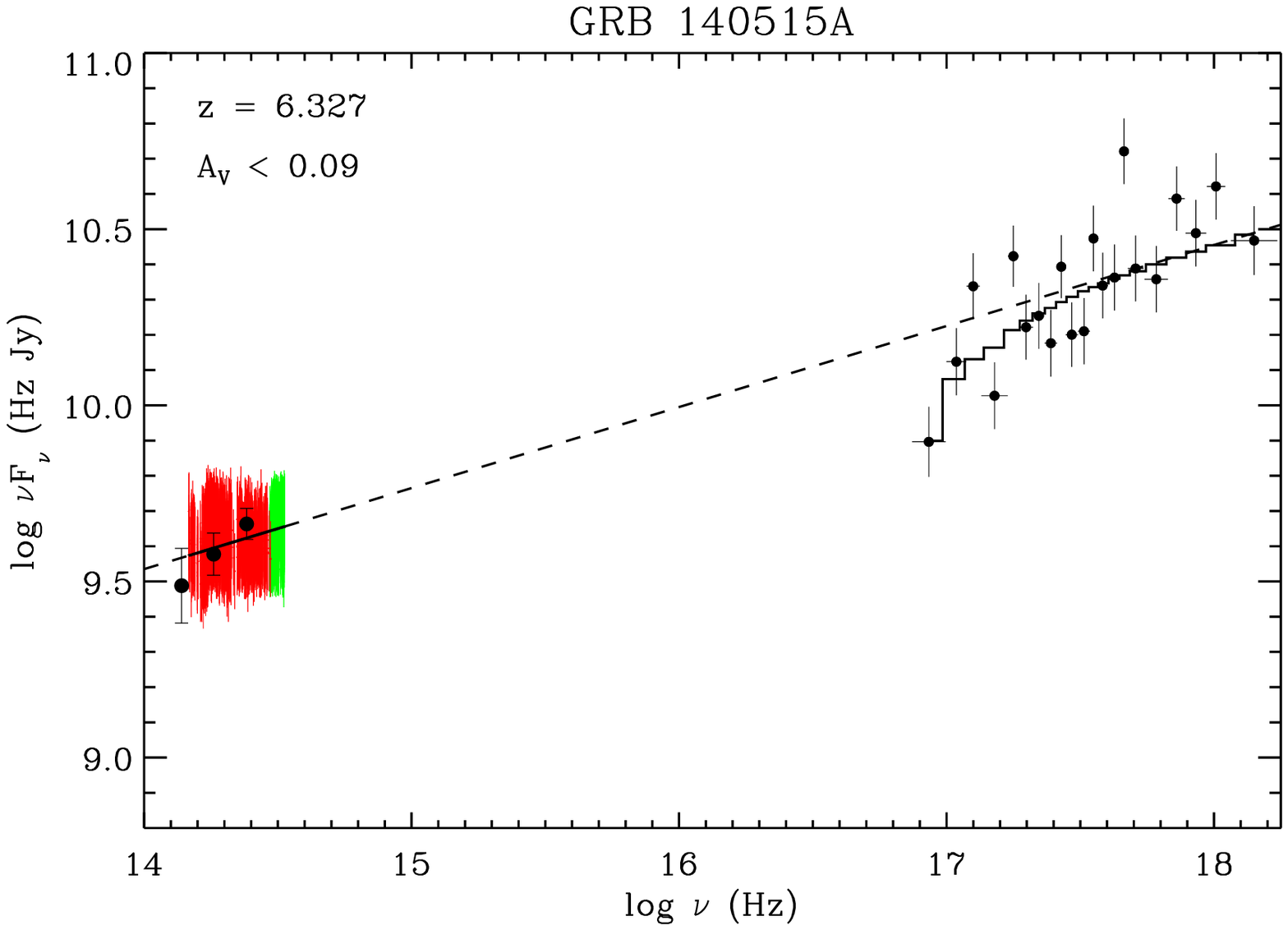}}
{\includegraphics[width=\columnwidth,clip=]{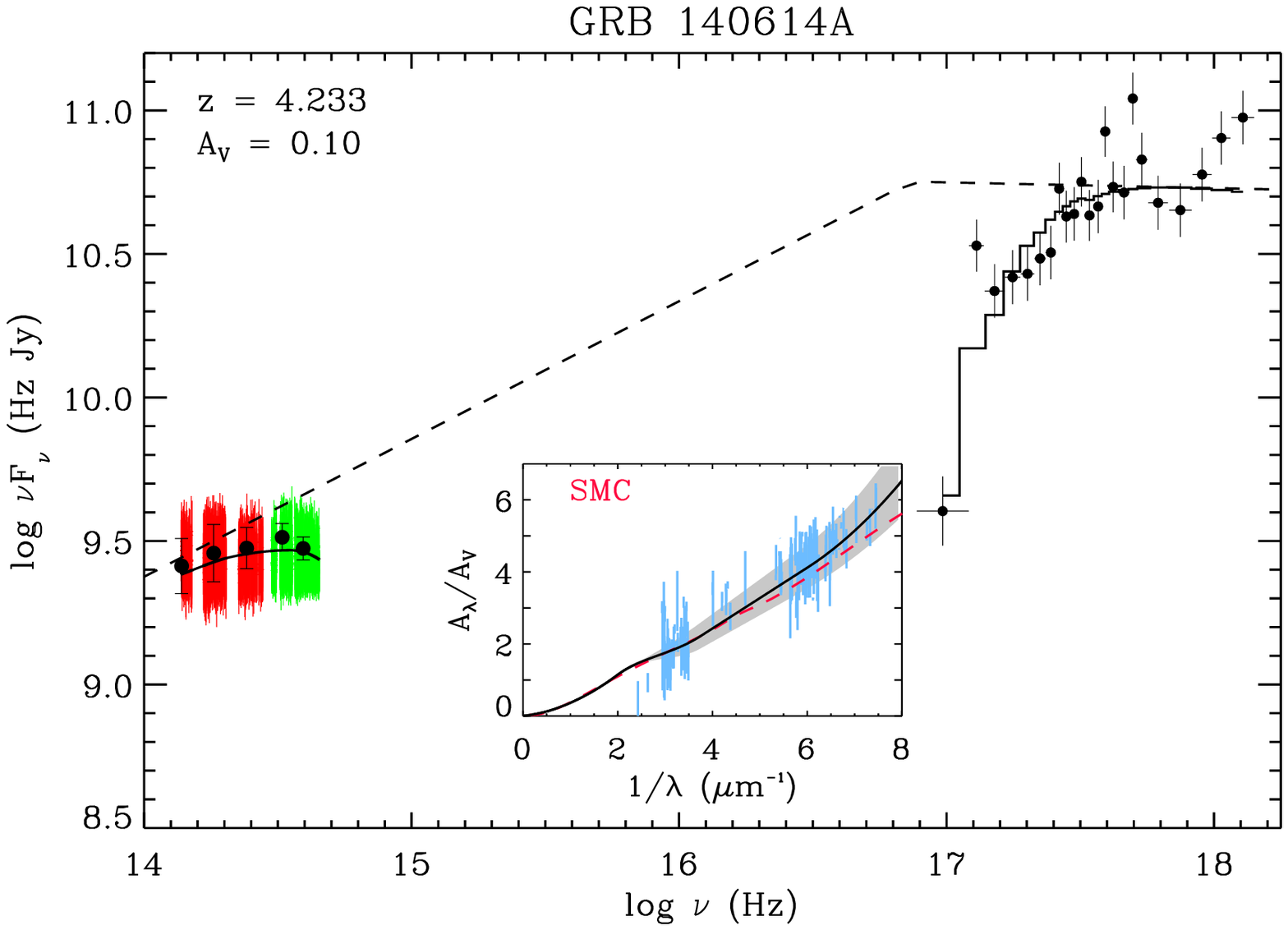}}
{\includegraphics[width=\columnwidth,clip=]{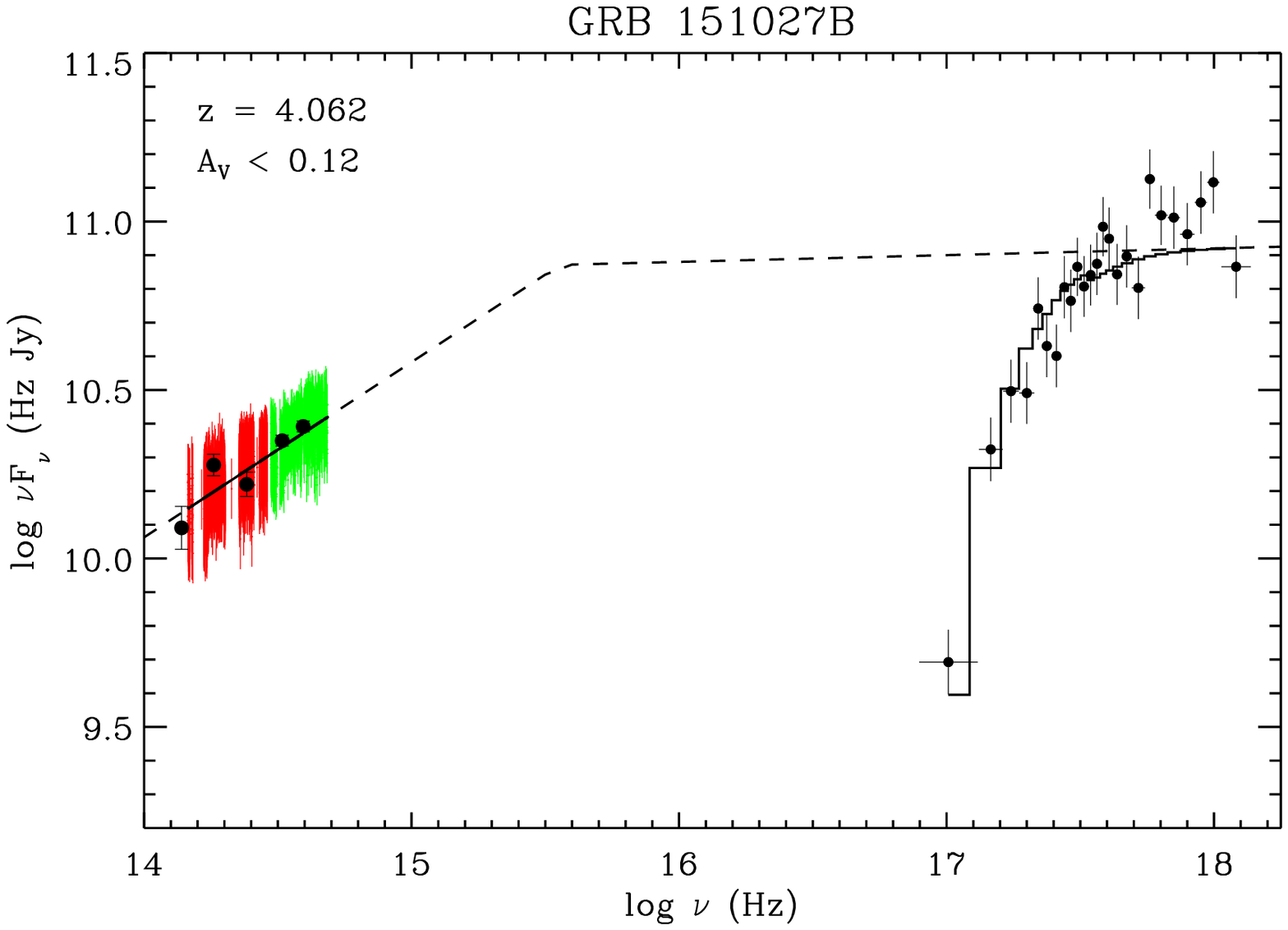}}
{\includegraphics[width=\columnwidth,clip=]{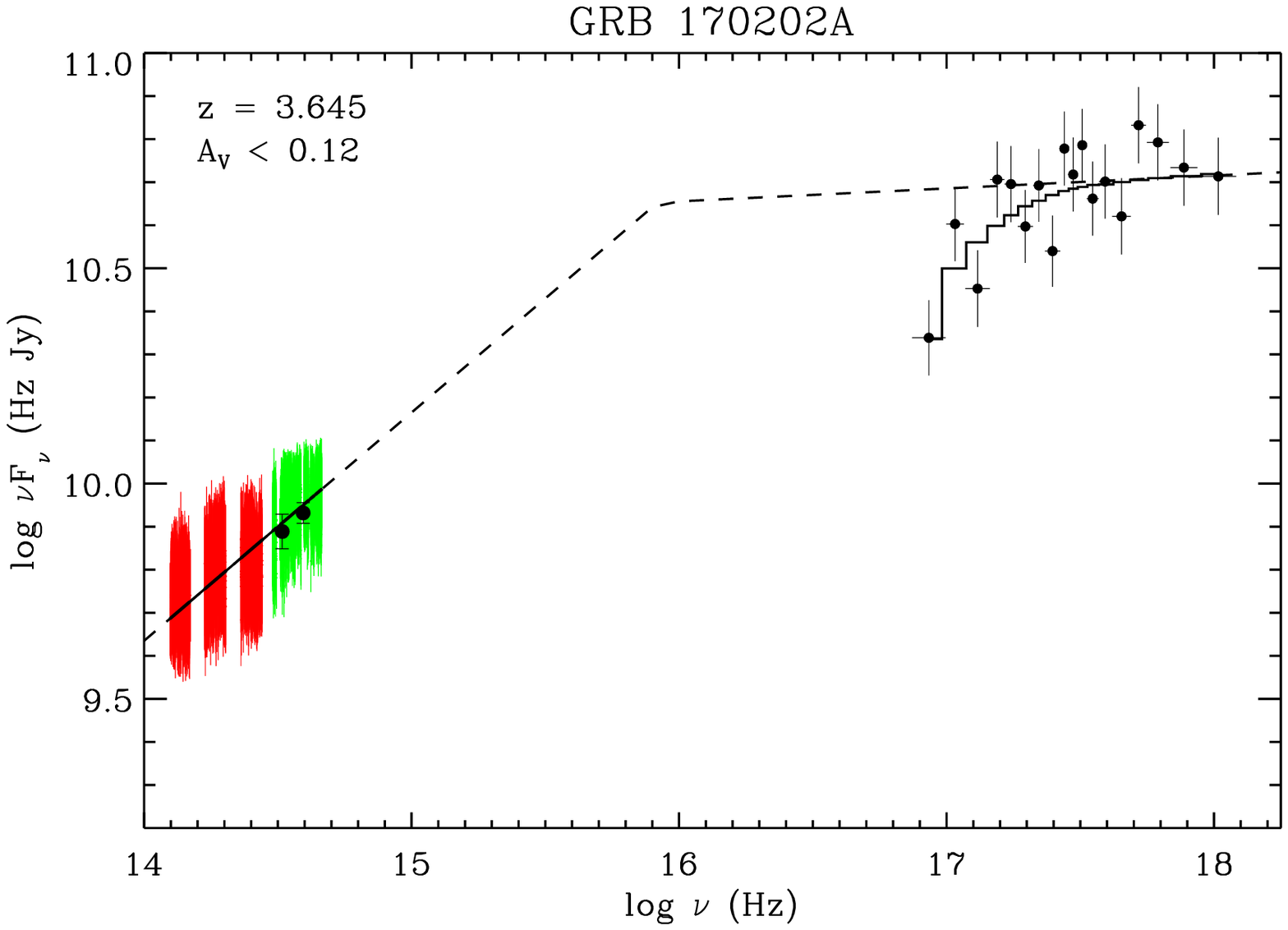}}
     \caption{Continued.}
  \end{figure*}

\subsection{X-shooter data}
The X-shooter spectra and available multi-band photometric data were corrected for the foreground Galactic extinction using the maps of \citet{schlafly11}. For each case, the Galactic extinction value is provided in Table \ref{grb:list} and the uncertainties on $E(B-V)_{\rm Gal}$ are very small and would have a negligible effect on our results. Only in the case of GRB\,151027B, the $E(B-V)_{\rm Gal}$ uncertainty is 0.03\,mag and could alter our results by 10\%\ but that case is consistent with no dust (see \S\ref{ebvdis}). The SEDs are generated at photometric mid-times, $\Delta$t, and the X-shooter spectra are scaled to the photometric observations. Usually the photometric and spectroscopic data are comparable, however, sometimes they differ up to 15\%. We used the HEAsoft software (version 6.19) tool \texttt{flx2xsp} to convert X-shooter data to the XSPEC (version12.9; \citealt{arnaud96}) readable spectral (PHA) and response matrices (RSP) files. The data around and below the damped Ly$\alpha$ absorber, metal absorption lines, atmospheric telluric absorption, and spikes originating from sky subtraction residuals were masked out. The masking file, PHA, and RSP files were grouped using the \texttt{grppha} tool and individual data channels (representing the spectral binning) were kept and no additional re-binning is applied.

\subsection{X-ray data}
The X-ray data for each GRB afterglow is obtained from the \emph{Swift} X-ray Telescope (XRT, \citealt{burrows05}). The XRT lightcurves were gathered from the \emph{Swift} online repository \citep{evans09} and a time decay model \citep{beuermann99} is fitted to the data. The X-ray spectrum for each GRB afterglow is reduced using the HEAsoft software in the 0.3--10.0\,keV energy range around the SED mid-time $\Delta t$ (which is the photometric midpoint). We used photon counting (PC) mode observations and selected data around the region of no spectral evolution. The X-ray data are extracted using the \texttt{XSELECT} (version 2.4) tool and RSP files were used from the \emph{Swift} XRT calibrations. The X-ray PHA and RSP files were grouped to 20 counts per energy channel using the \texttt{grppha} tool. This is done to have a better handle on the X-ray slopes. The X-ray lightcurves were then used to estimate a ratio of the SED mid-time $\Delta t$ and the photon weighted mean time of the X-ray spectrum was further applied to normalise the X-ray spectra.

\section{SED and dust fitting procedure}
 Theoretically, GRBs are defined by the `fireball model' \citep[e.g,][]{meszaros97}, suggesting GRB afterglows originate from synchrotron radiation caused by the interaction between the ultra-relativistic jet and the ISM. Cooling of electrons, in the GRB postshock, produces a break in the synchrotron spectrum. This cooling break is sometimes located between the optical and X-ray bands \citep[see Fig.1 of][]{sari98} with a well-defined change in spectral slope of $\Delta\beta=0.5$. This fixed change in the slope is supported by analysis of both spectroscopic \citep{zafar11} and photometric \citep{greiner11} GRB SEDs.
 
We followed the SED and dust fitting method described in \citet{zafar18}. We briefly outline the method here. For a full description we refer the reader to \citet{zafar18}. We used the spectral fitting package \texttt{XSPEC} to fit the restframe X-ray to the optical/NIR SEDs of $z\geq3$ GRB afterglows. A single or broken power-law together with a parametric extinction law is used to model the SEDs. In case of a single power-law, the intrinsic spectral shape is defined by the slope $\beta_{\rm opt}$. In case of a broken power law a cooling break ($\nu_{\rm break}$) is required. In those cases the intrinsic slopes, $\beta_{\rm opt}$ (optical slope) and $\beta_{\rm X}$ (X-ray slope), were fitted such that the change in slope ($\Delta\beta$) was fixed at 0.5 (\citealt{sari98}, see also \citealt{zafar11,greiner11}).

For the X-ray data, the total Galactic equivalent neutral hydrogen column density ($N_{\rm H, Gal}$) was fixed within \texttt{XSPEC} using $tbabs$ to the values estimated from \citet{willingale13}. \citet{willingale13} values include atomic hydrogen column density \citep{kalberla05} and contributions from molecular hydrogen \citep{wilms00} and Galactic dust \citep{schlegel98}. The soft X-ray absorption indicating the restframe host galaxy equivalent neutral hydrogen column density, $N_{\rm H,X}$, is left as a free parameter using $ztbabs$. We used the \texttt{XSPEC} default solar abundances of \citet{anders89} following the discussions of \citet{watson11} and \citet{watson13}.

The observed spectra are changed due to dust scattering and absorption and given as: $F_{\nu}^{\rm{obs}} = F_\nu10^{-0.4A_\lambda}$, where $A_\lambda$ is described by the \citet{fm90} dust model. The \citet{fm90} law provides more freedom in fitting the extinction curves using two components: $i)$ a UV linear component defined by $c_1$ (intercept) and $c_2$ (slope) parameters with $c_4$ providing the far-UV curvature with far-UV term $F(\lambda^{-1})$ and $ii)$ a Drude component specifying the 2175\,\AA\ extinction bump by $c_3$ (bump strength), $x_0$ (central wavelength), and $\gamma$ (bump width) parameters. The wavelength-dependent extinction, $A_\lambda$, is thus given as:
\begin{equation}
A_\lambda = \frac{A_V}{R_V}\times\left(c_1+c_2\lambda^{-1}+c_3D(x,x_0,\gamma) +c_4F(\lambda^{-1}) + 1\right)
\end{equation}
Where $F(\lambda^{-1})=0$ for $\lambda^{-1}<5.9$\,$\mu$m$^{-1}$ and $F(\lambda^{-1})=$ $0.5392(\lambda^{-1}-5.9)^2+0.05644(\lambda^{-1}-5.9)^3$ for $\lambda^{-1}\ge5.9$\,$\mu$m$^{-1}$. We initially fitted the data with the Drude component to search for a 2175\,\AA\ bump but in all cases, the bump strength was consistent with zero. Therefore, we fixed the Drude component to $c_3=0$, $\gamma=1$\,$\mu m^{-1}$, and $x_0=4.6$\,$\mu m^{-1}$. This was done to avoid degeneracies and to allow a better constraint for other parameters, following the discussions of \citet{zafar15}.

Finally, for the $z\geq3$ GRB SED analysis within \texttt{XSPEC}, $c_1$, $c_2$, $c_4$, $A_V$, $R_V$, ($N_{\rm H,X}$), $\nu_{\rm break}$, and spectral indices of the continuum ($\beta_{\rm opt}$ and $\beta_{\rm X}$), were fitted as free parameters. We refer the reader to \citet{zafar18} for discussions about correlated parameters and their errors. The broken power-law model is considered a better fit for the cases where the F-test probability is smaller than 5\%. In Table \ref{best-fit}, the best fit results for both single and broken power-laws, resulting reduced $\chi^2$ (derived using errors in the optical and X-ray spectra), and the null hypothesis probability for each case are provided.

\section{Results}\label{grb:det}
In this section, we provide details of the spectroscopic and photometric data collection and SED construction for each case. The SED fitting is performed on the X-shooter data and XRT data. The photometric data are not included in the SED analysis and only used for correcting the sub-optimal flux-calibration of X-shooter. The best-fit models of each GRB afterglow and extinction curves for the dusty cases are shown in Fig. \ref{grb:fits}.

\subsection{GRB\,110818A}
The X-shooter spectra of the GRB\,110818A ($z=3.360$) afterglow were carried out at $\sim$6.2\,hours after the burst trigger. The only available photometric measurement in the $R$-band is from the X-shooter acquisition camera \citep{avanzo11}. We used the $R$-band observation for the X-shooter-XRT SED normalisation. The SED prefers a broken power-law and a featureless extinction curve ($R_V=2.61^{+0.14}_{-0.15}$) with $A_V=0.23^{+0.03}_{-0.04}$\,mag.

\subsection{GRB\,111123A}
The X-shooter spectra of the GRB\,111123A ($z=3.152$) afterglow were taken at $\sim$13.92\,hours after the burst trigger. The afterglow was detected by GROND only in the $r^{\prime}$ band and was undetected in the remaining bands \citep[see also][]{rossi11}. We used that $r^{\prime}$ band data for the X-shooter-XRT SED normalisation. The SED prefers a broken power-law and no dust extinction with $A_V<0.14$\,mag.

\subsection{GRB\,120712A}
The X-shooter spectra of GRB\,120712A ($z=4.1745$) were obtained at $\sim$11.04\,hours after the burst trigger. Photometric data are obtained from \citet{bolmer17} in the $i^{\prime}z^{\prime}JH$ and $K$ bands from GROND at 10.43\, hours after the burst. The $i^{\prime}$ and $H$-band data are used to scale the VIS and NIR arm spectra, respectively. The SED is fit well with a broken power-law and an SMC-like extinction curve ($R_V=2.73^{+0.18}_{-0.23}$) with $A_V=0.08\pm0.03$\,mag. Due to the shape of the photometric data, we also attempted to fit the GROND-XRT SED using a fixed \citet{pei92} SMC law. The photometric SED is fit well with a broken power-law and $A_V=0.05^{+0.03}_{-0.04}$\,mag. Previously, \citet{bolmer17} find that the photometric SED at 10.6\,hrs is fit well with a broken power-law and an SMC curve with $A_V=0.08^{+0.03}_{-0.08}$\,mag, consistent with our findings.

\begin{table*}
\caption{Results of the best-fit parameters of $z\geq3$ GRB SEDs. For each GRB single (first row) and broken power-law (second row) results of fits to the SEDs are provided. The columns indicate: 1) the burst name, 2) the equivalent neutral hydrogen column density ($N_{\rm H,X}$), 3) optical slope ($\beta_{\rm opt}$), 4) X-ray slope ($\beta_{\rm X}$), 5) cooling break frequency ($\nu_{\rm break}$), 6) UV intercept $c_1$, 7) UV slope $c_2$, 8) far-UV curvature $c_4$, 9) total-to-selective extinction $R_V$, 10) visual extinction $A_V$, (11) reduced $\chi^2$ with number of degrees of freedom (dof) and 12) the Null Hypothesis Probability (NHP). The best-fit models are denoted by $\dagger$. The second-last row provides the weighted mean ($WM$) values and 1$\sigma$ errors (error on $WM$) of all best-fit extinction curves parameters. The standard deviations (intrinsic scatter) around the $WM$ values are provided in the last row.}
\label{best-fit}
\centering
\renewcommand{\footnoterule}{}  
\setlength{\tabcolsep}{1.5pt}
\begin{tabular}{l c c c c c c c c c c c}\hline\hline
GRB & $N_{\rm H,X}$ & $\beta_{\rm opt}$ & $\beta_{\rm X}$ & log $\nu_{\rm break}$ & $c_1$ & $c_2$ & $c_4$ & $R_V$ & $A_V$ & $\chi^2_\nu$/dof  & NHP\% \\
	& 10$^{22}$ cm$^{-2}$ &  & & Hz & $\mu$m &  $\mu$m$^2$ 	& & & mag & & prob. \\
\hline\hline
110818A & $<1.72$ & $0.71^{+0.14}_{-0.15}$ & $\cdots$ & $\cdots$ & $-4.27\pm0.16$ & $1.98\pm0.13$ & $0.41\pm0.11$ & $3.77^{+0.21}_{-0.24}$ & $0.19\pm0.05$ & 1.02/34827 & 0.40 \\ 
 & $1.84^{+0.95}_{-0.76}$ & $0.48^{+0.10}_{-0.08}$ & $0.98^{+0.12}_{-0.13}$ & $15.88\pm0.17$ & $-4.75\pm0.15$ & $2.24\pm0.10$ & $0.43\pm0.09$ & $2.61^{+0.14}_{-0.15}$ & $0.23^{+0.03}_{-0.04}$ & 0.98/34825 & 100$^\dagger$ \\
111123A & $<0.38$ & $0.52^{+0.15}_{-0.12}$  & $\cdots$  & $\cdots$ & $\cdots$ & $\cdots$ & $\cdots$ & $\cdots$ & $<0.12$ & 0.99/37065 & 91.0 \\
& $<1.92$ & 0.41$^{+0.13}_{-0.12}$  & $0.91^{+0.12}_{-0.14}$  & $16.93\pm0.17$ & $\cdots$ & $\cdots$ & $\cdots$ & $\cdots$ & $<0.14$ & 0.99/37063 & 97.0$^\dagger$ \\
120712A & $<1.97$ & $1.10^{+0.11}_{-0.16}$ & $\cdots$ & $\cdots$ & $2.80\pm0.17$ & $-0.60\pm0.16$ & $0.09\pm0.12$ & $4.56^{+0.38}_{-0.41}$ & $0.90^{+0.12}_{-0.15}$ & 1.24/29055 & $0.00$ \\
 & $<2.35$ & $0.64^{+0.14}_{-0.15}$ & $1.14^{+0.08}_{-0.13}$ & $15.15\pm0.21$ & $-4.93\pm0.12$ & $2.28\pm0.11$ & $0.47\pm0.09$ & $2.73^{+0.18}_{-0.23}$ & $0.08^{+0.03}_{-0.03}$ & 0.99/29053 & 96.0$^\dagger$ \\
121201A & $<0.61$ & $0.75^{+0.12}_{-0.14}$ & $\cdots$ & $\cdots$ & $-4.62\pm0.19$ & $2.12\pm0.14$ & $0.72\pm0.14$ & $2.42^{+0.21}_{-0.20}$ & $0.10\pm0.03$ & 1.00/32050 & $42.0$ \\
 & $<1.59$ & $0.52^{+0.12}_{-0.12}$ & $1.02^{+0.11}_{-0.09}$ & $16.03\pm0.13$ & $-5.06\pm0.17$ & $2.21\pm0.13$ & $0.91\pm0.10$ & $2.53^{+0.22}_{-0.19}$ & $0.09^{+0.02}_{-0.03}$ & 0.95/32048 & 100$^\dagger$ \\
130408A & $<0.36$ & $0.68^{+0.11}_{-0.14}$ & $\cdots$ & $\cdots$ & $-4.71\pm0.15$ & $1.95\pm0.12$ & $0.29\pm0.12$ & $2.71^{+0.16}_{-0.15}$ & $0.20^{+0.05}_{-0.04}$ & 1.01/26868 & 23.0 \\ 
 & $<0.40$ & $0.55^{+0.09}_{-0.12}$ & $1.05^{+0.08}_{-0.10}$ & $16.82\pm0.08$ & $-4.92\pm0.13$ & $2.26\pm0.10$ & $0.33\pm0.14$ & $2.83^{+0.14}_{-0.17}$ & $0.22^{+0.04}_{-0.05}$ & 1.01/26866 & 26.0$^\dagger$\\ 
140311A & $<2.25$ & $0.86^{+0.10}_{-0.11}$ & $\cdots$ & $\cdots$ & $-4.77\pm0.15$ & $2.19\pm0.13$ & $0.43\pm0.14$ & $2.66^{+0.17}_{-0.19}$ & $0.15^{+0.05}_{-0.04}$ & 0.75/15536 & 100$^\dagger$ \\
  & $<4.16$ & $0.60^{+0.13}_{-0.11}$ & $1.10^{+0.12}_{-0.12}$ & $15.85\pm0.10$ & $-3.93\pm0.16$ & $1.94\pm0.12$ & $0.34\pm0.16$ & $3.21^{+0.15}_{-0.16}$ & $0.23^{+0.03}_{-0.05}$ & 1.21/15538 & 0.00 \\
140515A & $<3.78$ & $0.77^{+0.15}_{-0.13}$ & $\cdots$ & $\cdots$ & $\cdots$ & $\cdots$ & $\cdots$ & $\cdots$ & $<0.09$ & 0.97/19553& 100$^\dagger$ \\
  & $<6.24$ & $0.62^{+0.13}_{-0.16}$ & $1.12^{+0.14}_{-0.15}$ & $16.86\pm0.17$ & $\cdots$ & $\cdots$ & $\cdots$ & $\cdots$ & $<0.10$ & 1.14/19555 & 0.00 \\
140614A & $<3.14$ & $0.63^{+0.13}_{-0.12}$ & $\cdots$ & $\cdots$ & $-4.61\pm0.15$ & $2.21\pm0.16$ & $0.42\pm0.12$ & $2.73^{+0.16}_{-0.21}$ & $0.07^{+0.02}_{-0.04}$ & 1.02/20516 & $2.25$ \\
  & $<4.89$ & $0.52^{+0.10}_{-0.09}$ & $1.02^{+0.07}_{-0.10}$ & $16.87\pm0.10$ & $-5.03\pm0.18$ & $2.18\pm0.14$ & $0.47\pm0.13$ & $2.59^{+0.19}_{-0.17}$ & $0.10^{+0.04}_{-0.03}$ & 0.93/20514 & 100$^\dagger$ \\
151027B & $<38.22$ & $0.52^{+0.14}_{-0.16}$ & $\cdots$ & $\cdots$ & $\cdots$ & $\cdots$ & $\cdots$ & $\cdots$ & $<0.11$ & 1.07/28725 & $0.00$ \\
 & $<7.93$ & $0.48^{+0.08}_{-0.11}$ & $0.98^{+0.09}_{-0.06}$ & $15.56\pm0.14$ & $\cdots$ & $\cdots$ & $\cdots$ & $\cdots$ & $<0.12$ & 1.01/28723 & $12.0$$^\dagger$ \\
170202A & $<7.26$ & $0.65^{+0.11}_{-0.16}$ & $\cdots$ & $\cdots$ & $\cdots$ & $\cdots$ & $\cdots$ & $\cdots$ & $<0.13$ & 1.06/26322 & $0.00$ \\
 & $<1.31$ & $0.47^{+0.11}_{-0.14}$ & $0.97^{+0.15}_{-0.10}$ & $15.92\pm0.16$ & $\cdots$ & $\cdots$ & $\cdots$ & $\cdots$ & $<0.12$ & 1.02/26320 & $2.00$$^\dagger$ \\
\hline
$WM$ & $\cdots$ & $\cdots$ & $\cdots$ & $\cdots$ & $-4.90\pm0.07$ & $2.24\pm0.06$ & $0.53\pm0.04$ & $2.67\pm0.08$ & $\cdots$ & $\cdots$ \\
$Stddev$ & $\cdots$ & $\cdots$ & $\cdots$ & $\cdots$ & $0.21$ & $0.14$ & $0.05$ & $0.22$ & $\cdots$ & $\cdots$ \\
\hline
\end{tabular}
\end{table*}


\subsection{GRB\,121201A}
The X-shooter spectra of GRB\,121201A ($z=3.385$) were obtained at $\sim$12.87\,hours after the burst trigger. Photometric data were taken with GROND in the $r^\prime i^\prime$ and $z^\prime$ bands from \citet{kruhler12}. We used the $i^\prime$-band magnitude for the SED normalisation. The X-ray to the NIR SED is fit well with a broken power-law and and a featureless extinction curve ($R_V=2.53^{+0.22}_{-0.18}$) with $A_V=0.09^{+0.02}_{-0.04}$\,mag.

\subsection{GRB\,130408A}
The X-shooter spectra of the afterglow of GRB\,130408A ($z=3.758$) were taken at $\sim2$\,hours after the burst. Photometric data from GROND are reported at 1.5\,hours after the burst \citep{sudilovsky13} in the $i^\prime z^\prime JH$ and $K$ bands. We used the $i^{\prime}$ and $H$-band data to normalise the VIS and NIR arm spectra, respectively. The SED provides a good fit with a broken power-law and a featureless extinction curve ($R_V=2.83^{+0.14}_{-0.17}$) with $A_V=0.22^{+0.04}_{-0.05}$\,mag. The GROND-XRT SED alone suggests a broken power-law with $A_V=0.20\pm0.06$\,mag (using a fixed \citealt{pei92} SMC extinction curve) describe the data well. Previously, \citet{wiseman17} reported $A_V=0.22\pm0.03$\,mag for this burst, consistent with our results.

\subsection{GRB\,140311A}
The X-shooter spectra of GRB\,140311A ($z=4.9545$) were taken at 32.5\,hours after the burst trigger. Nordic Optical Telescope (NOT) photometry in the $i^\prime$ band is available at 27.89\,hours after the burst from \citet{malesani14}. We normalised the X-shooter spectra to the level of the NOT photometry to fit the intrinsic extinction curve. The SED prefers a single power-law and a featureless steep extinction curve ($R_V=2.66^{+0.17}_{-0.19}$) with $A_V=0.15^{+0.05}_{-0.04}$\,mag. \citet{bolmer17} found that GROND-XRT data at 9.8\,hours are fit well with a single power-law and an SMC extinction curve with $A_V=0.07\pm0.03$\,mag, consistent within 2$\sigma$. However, \citet{laskar17} found that the X-ray to optical SED is best described by the extinction of $A_V\approx0.3$\,mag at 0.4\,days after the burst.

\subsection{GRB\,140515A}
The spectra of the highest redshift burst ($z=6.327$) of our sample were taken with X-shooter at $\sim16.32$\,hours after the burst trigger. At 16.83\,hours after the burst, photometric data in the $JH$ and $K_s$ bands were taken with GROND by \citet{bolmer17}. We normalised the X-shooter observations to the $H$-band photometric data. The SED at 16.83\,hours after the burst is fit well with a single power-law and no extinction with $A_V<0.09$\,mag. The GROND-XRT SED also suggests a best-fit with a single power-law and $A_V<0.10$\,mag. \citet{bolmer17} reported no dust extinction with $A_V<0.1$\,mag from the GROND-XRT analysis at 14.6\,hours after the burst, consistent with our findings. Previously, \citet{melandri15} found the SED of this burst prefers an SMC-type extinction curve with a small amount of extinction of $A_V=0.11\pm0.02$\,mag. However, \citet{mcguire16} found no dust for this burst with $A_V\leq0.1$\,mag.

\subsection{GRB\,140614A}
The X-shooter spectra of GRB\,140614A ($z=4.233$) were obtained at $\sim$3.9\,hours after the burst trigger. GROND photometric data were obtained from \citet{bolmer17} in the $i^\prime z^\prime JH$ and $K$ bands at 3.15\,hours after the burst. We used the $i^{\prime}$ and $H$-band data to scale the VIS and NIR arm spectra, respectively. The X-shooter and X-ray spectra are scaled to the photometry. The SED is fit well with a broken power-law and a featureless SMC-like extinction curve ($R_V=2.59^{+0.19}_{-0.17}$) with $A_V=0.10^{+0.04}_{-0.03}$\,mag. The GROND-XRT SED fit well with a broken power-law and $A_V<0.15$\,mag. Previously \citet{bolmer17} found the best fit with a broken power-law and an SMC-type extinction curve with $A_V=0.11^{+0.17}_{-0.05}$\,mag for this burst at 4.1\,hours after the burst, suggesting consistent results.

\subsection{GRB\,151027B}\label{ebvdis}
The X-shooter spectra of GRB\,151027B ($z=4.062$) were acquired at $\sim$5.4\,hours after the burst. Photometric data at 8.01\,hours were obtained with GROND in the $i^\prime z^\prime JH$ and $K$ bands \citep{bolmer17}. We used the $i^{\prime}$ and $J$-band photometry to normalise the VIS and NIR arm spectra, respectively. The SED is fit well with a broken power-law and no dust extinction with $A_V<0.12$\,mag. The GROND-XRT SED is also fit well with a broken power-law and $A_V<0.07$\,mag. Previously, \citet{bolmer17} found the SED at 8.8\, hours for this burst to be well fitted a broken power-law and no extinction with $A_V<0.2$\,mag, suggesting consistent results. Recently, \citet{greiner18} reported that the SED of this burst could be well-explained by a single power-law and a negligible amount of dust with $A_V<0.04$\,mag.



\subsection{GRB\,170202A}
The X-shooter spectra of the GRB\,170202A ($z=3.645$) afterglow were taken at $\sim$9.7\,hours after the burst trigger. Suitable photometric observations in the $i^\prime$ and $z^\prime$ bands were taken from the 2.1m Otto Struve telescope at the McDonald Observatory at 16 hours after the burst \citep{im17}. The $i^\prime$ band photometry is used for the X-shooter spectra normalisation. The SED prefers a broken power-law and no dust extinction with $A_V<0.12$\,mag.

\subsection{X-ray analysis}
\begin{figure}
  \centering
{\includegraphics[width=\columnwidth,clip=]{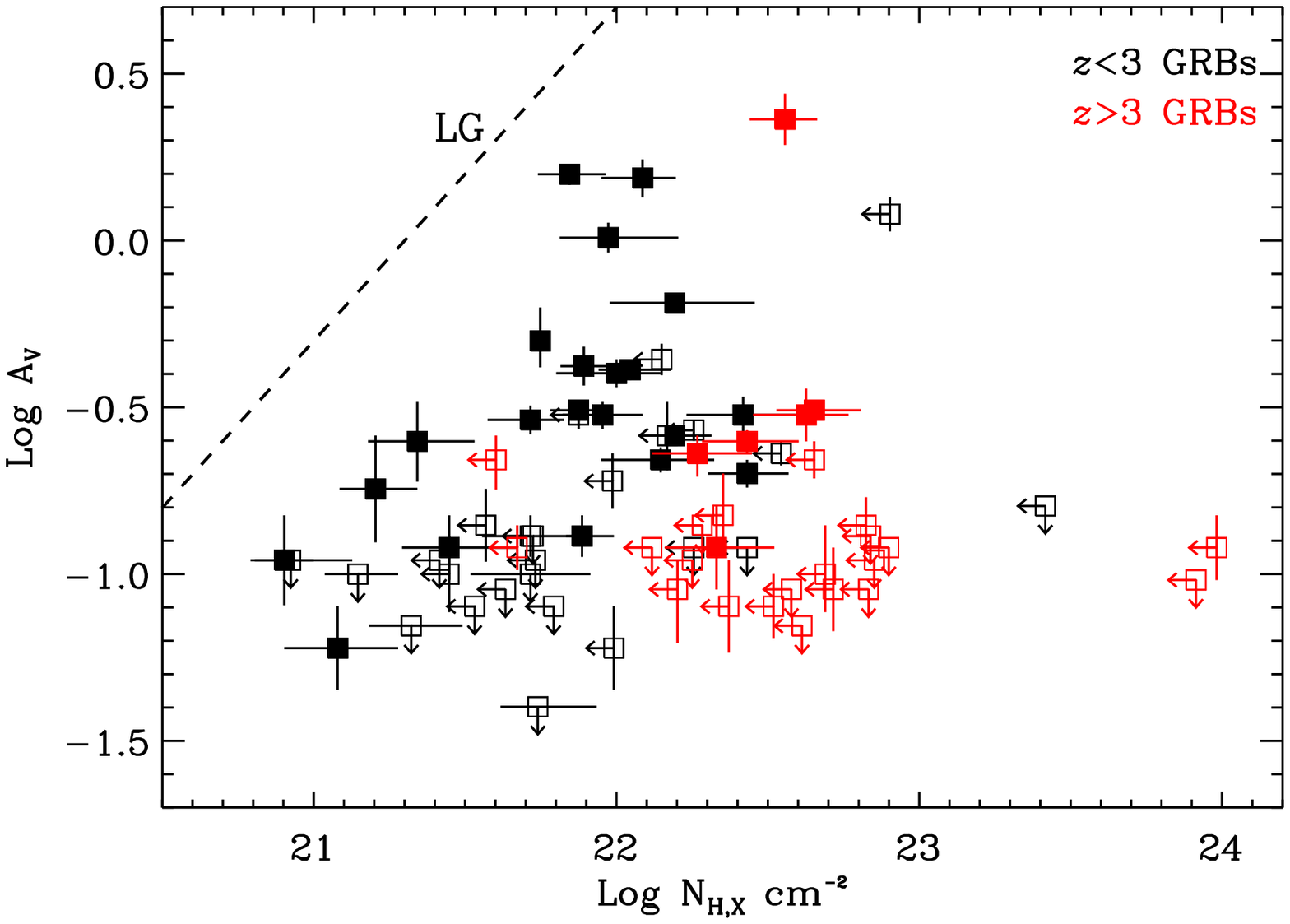}}
     \caption{Visual extinction against $N_{\rm H,X}$ for the spectroscopic GRBs where  SED fitting includes X-ray to optical/NIR data. The black points correspond to $z<3$ GRBs. The open symbols represent limits and filled symbols indicate measurements. The dashed curve represents the metals-to-dust ratio for the Local Group environments (indicated as LG).}
         \label{nhxav}
           \end{figure}

The equivalent hydrogen column densities ($N_{\rm H,X}$) derived through the simultaneous X-shooter to X-ray SED fitting for each GRB case are reported in Table \ref{best-fit}. We obtained a significant $N_{\rm H,X}$ measurement for a single burst in our sample. We combined our $N_{\rm H,X}$ and $A_V$ results with other values derived through the spectroscopic optical/NIR to X-ray SED analyses and reported in \citet{zafar11,zafar11b,zafar18}. We split the $N_{\rm H,X}$ data in redshift  below and above $z=3$. The results indicate an increase in the equivalent hydrogen column densities at $z\geq3$. Previously, \citet{campana12,starling13,campana15} suggested an evolution of $N_{\rm H,X}$ with increasing redshift. This increase is interpreted to be due to the growing intergalactic medium absorption at larger distances. In contrast, \citet{buchner17} claimed no evolution of $N_{\rm H,X}$ with redshift and find that the $N_{\rm H,X}$ distribution is an axisymmetric ellipsoid of gas having  randomly distributed sources within. Fig. \ref{nhxav} shows the $N_{\rm H,X}/A_V$ variation below and above $z=3$. Such an evolution of $N_{\rm H,X}/A_V$ with redshift has been found by \citet{watson13}, suggesting helium in \hii\ regions or metals ejected by the star could be the dominant X-ray absorber. The increase in $N_{\rm H,X}$ could also simply be explained by a larger gas column density in GRB hosts at higher redshifts \citep{heintz18}.


\section{Discussion}
\subsection{$z\geq3$ GRB extinction curve}
\begin{figure}
  \centering
{\includegraphics[width=\columnwidth,clip=]{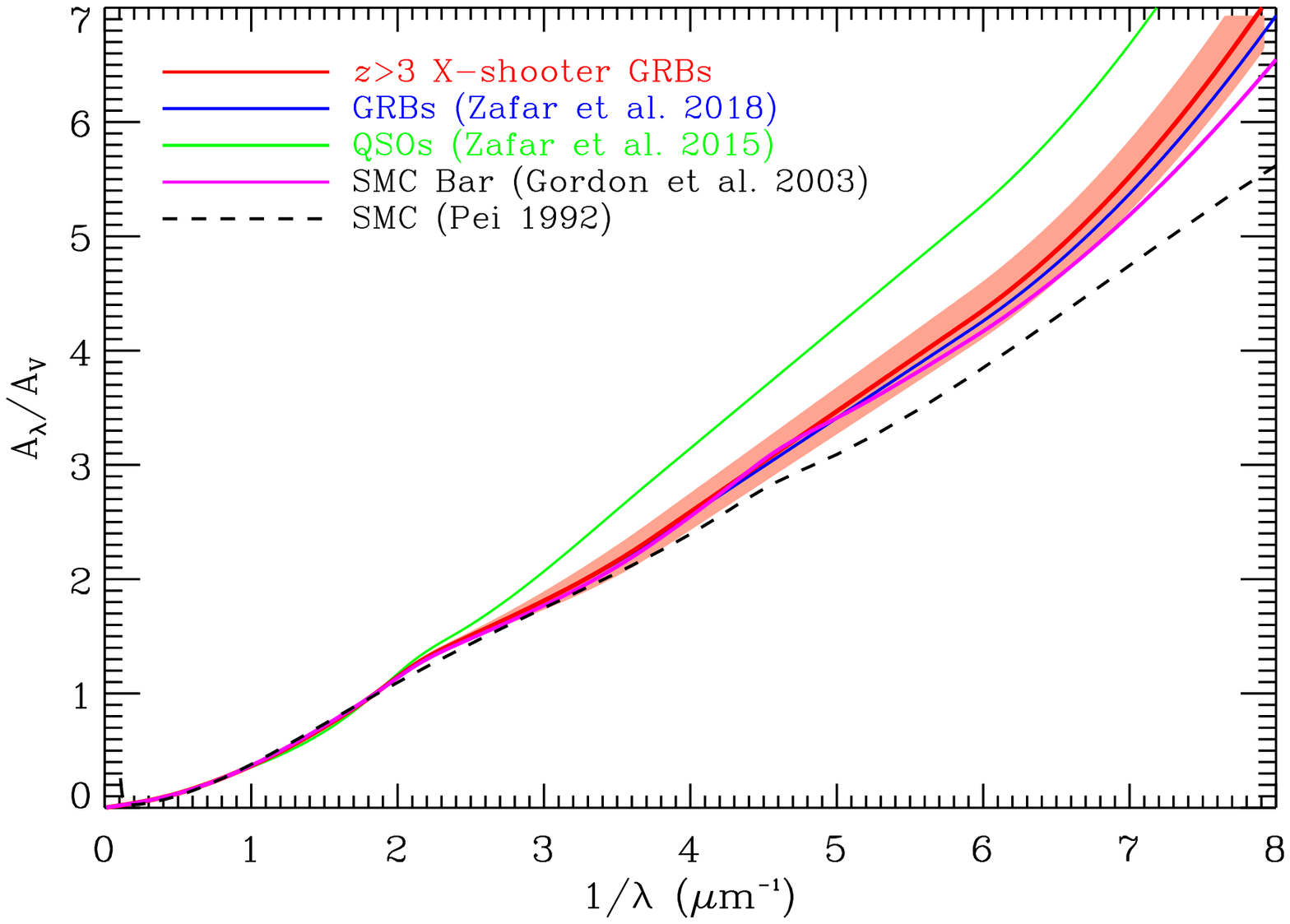}}
     \caption{$z\geq3$ GRB extinction curve (red) from the X-shooter data using our sample and $z\geq3$ GRBs from \citet{zafar18}. The red shaded area represents the 1$\sigma$ error region of the extinction curve parameters. For a comparison other featureless extinction curves from GRBs \citep[blue;][]{zafar18}, QSOs \citep[green;][]{zafar15}, SMC Bar \citep[magenta;][]{gordon03}, and the typical SMC \citep[dashed black;][]{pei92} are plotted.}
         \label{ext:curves}
           \end{figure}

Six GRBs in our sample are extinguished and their best-fit extinction parameters are provided in Table \ref{best-fit}. We calculated the {\it weighted mean} ($WM$) values and standard deviations of our extinction curve parameters (see last row of Table \ref{best-fit}). The $WM$ $R_V$ of our $z\geq3$ sample is found to be $R_V=2.67$ (with a scatter of 0.22), consistent with the value of $R_V\sim2.61\pm0.08$ by \citet{zafar18} for bursts at all redshifts. We combined our six best-fit extinction curve results with the three dusty $z\geq3$ X-shooter GRBs (GRB\,090313, GRB\,100219A, and GRB\,111008A) presented in the sample of \citet{zafar18}. This is done to generate an intrinsic extinction curve for $z\geq3$ X-shooter GRBs, resulting in $c_1=-5.03\pm0.04$, $c_2=2.29\pm0.05$, and $c_4=0.63\pm0.03$ and $R_V=2.60\pm0.06$. The resultant extinction curve for all $z\geq3$ X-shooter selected GRB afterglows is shown in Fig. \ref{ext:curves} with its 1$\sigma$ error as shaded area. The extinction curve is further compared with the previously selected X-shooter GRB sample \citep{zafar18}, a sub-sample of dusty quasars (QSOs; \citealt{zafar15}) found through the High $A_V$ quasar (HAQ; \citealt{krogager15}) survey, SMC Bar \citep{gordon03} and the typically used SMC-type \citet{pei92} extinction curves. Using the Kolmogorov-Smirnov statistics, the $z\geq3$ GRB extinction curve deviates from the canonical SMC-type curve at $>99$\% confidence level but is consistent at $\ge80$\% confidence level with the SMC-Bar extinction curve. For this reason, we suggest to use the SMC Bar extinction curve \citep{gordon03} rather than the \citet{pei92} SMC law for fitting the data with featureless extinction curves.

\subsection{Dust at high-redshift}\label{dustz}
\begin{figure}
  \centering
{\includegraphics[width=\columnwidth,clip=]{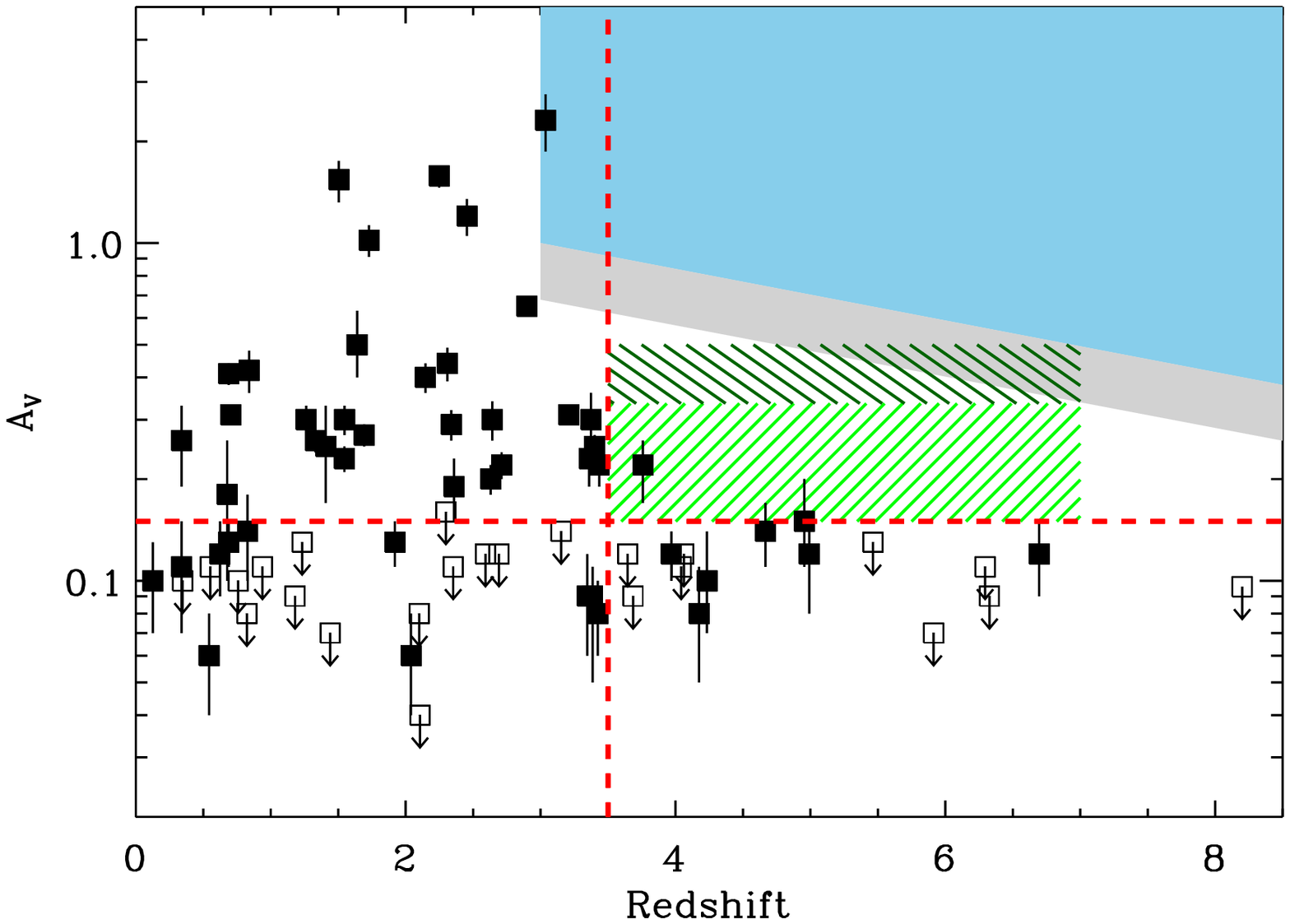}}
     \caption{Visual extinction in spectroscopic GRB afterglows against redshift. Combining our results with other spectroscopic GRB samples, we have 45 GRBs at $z<3$ and 27 GRBs at $z\geq3$. For illustration, red vertical and horizontal lines indicate $z=3.5$ and $A_V=0.15$\,mag marks, respectively. The gray and blue shaded regions represent the detection threshold for steeper and flatter optical slope bursts, respectively. The green shaded regions illustrate the areas from $A_V=0.15$ up to the detection threshold for both templates in $z=3.5-7$ range. }
         \label{zhist}
           \end{figure}

The restframe visual extinction for seven dusty GRBs in our sample ranges from $0.08^{+0.03}_{-0.03}$ to $0.23^{+0.03}_{-0.04}$\,mag. The remaining five are consistent with no extinction within their 3$\sigma$ $A_V$ limits given in Table \ref{best-fit}. We combined our $z\ge3$ sample with the spectroscopic GRB samples of \citet{zafar11,zafar11b,zafar18} and \citet{zafar18b}. This was done to be consistent throughout to compare samples where dust content is estimated using the spectroscopic SEDs. In Fig. \ref{zhist} we plot the visual extinction of this `full redshift coverage' sample versus redshift. It is seen that there appears to be an absence of GRB afterglows in the upper right quarter of that figure, i.e. an under-density of sighlines with moderate and higher ($A_V>0.15$\,mag) visual extinction at $z > 3.5$. A similar dearth of high $A_V$ detections at $z>4$ was previously reported by \citet{zafar11b}, \citet{bolmer17}. It is clear, however, that both high redshift and extinction will work towards making objects fainter and therefore more difficult to detect. To determine if the distribution of objects in Fig. \ref{zhist} can be fully explained by the two dimming effects we proceeded as follows.

We selected the spectra of GRB\,140311A and GRB\,140614A as templates to bracket our sample. The former has a steep optical slope ($\beta_{\rm opt}\sim0.9$) and a single power-law while the latter has a flatter slope ($\beta_{\rm opt}\sim0.5$) and a broken power-law. We then used the online VLT/X-shooter exposure time calculator (ETC) to calculate the X-shooter detection thresholds. Specifically we calculated the limiting magnitudes for one hour of observation with the same setup as used for the X-shooter ToO GRBs, and found that we would reach magnitudes of 23 and 22 in X-shooter bands $i^\prime$ and H, respectively. For each of the two template spectra we then redshifted them to a set of redshifts in the range 3 $<$ $z$ $<$ 8.5, and for each redshift determined the $A_V$ which would dim the target to the detection threshold. The resulting limits on detectability in the two cases are represented by the two differently shaded (blue and grey) regions in Fig. \ref{zhist}. We find that a moderately bright burst, such as the template of GRB\,140311A, is still detectable with an $A_V\sim1$\,mag at $z = 3$ and with an $A_V\sim0.4$\,mag back to $z = 8.5$. However, for the somewhat fainter burst GRB\,140614A an $A_V$ of 0.7 and 0.3\,mag will form the detection limits at $z=3$ and 8.5, respectively. In other words, for the pessimistic case with objects like GRB\,140614A we would expect to find objects in the trapeze shaped area above and to the right of the dashed red lines, but below the lower limit of the grey bar. For the optimistic case (GRB\,140311A) we would expect to find objects all the way up to the lower edge of the blue shaded area. To get conservative estimates of the expected number of afterglows which should have been detected in each of the two cases we consider now only the box-shaped hashed areas in Fig. \ref{zhist}. The green hashed box makes up about 60\% of the area below the lowest limit, the green+grey hashed box  makes up about 60\% of the area below the blue limit. For a simple prediction of the expected number we simply scale from the three other quarters. The total number of bursts with $A_V$ below 0.15\,mag and $z<7$ is 40, 15 of those found above $z>3.5$. In other words, in the right box we have 0.6 times the bursts in the left. We shall assume the same ratio (intrinsically) above $A_V = 0.15$\,mag. In the $A_V$ range corresponding to the green box (between 0.15 and 0.34) we have 20 below $z=3.5$ and therefore expect 20$\times$0.6 = 12 in the green box. In the more optimistic case of the larger box ($A_V$ between 0.15 and 0.50\,mag) we expect 25$\times$0.6 = 15. In both cases we observe only a single burst.  Our simple assumption of a constant $A_V$ distribution is therefore rejected at high significance, even in the very conservative case.

Spectroscopic follow-ups are carried out usually on the basis of $R$-band observations and the IGM drop occurring in the $R$-band for $z>4$ bursts leads to many cases not being spectroscopically observed. The number of detected high redshift GRBs will increase with the forthcoming Space Variable Object Monitor (SVOM) mission \citep{wei16}. Note that GRB\,080607 \citep{prochaska09} is one of the extremely bright cases detected at $z\sim3$ with $A_V\sim2.3$\,mag. Our study is also biased towards having simultaneous spectroscopic and photometric observations. Still a uniformly low dust content indicates a real decrease. Such a real decrease in dust content is a clue towards a change in dust properties at $z > 3.5$, but could also in part be caused by a not well-understood (though likely small) bias. \citet{bolmer17} using GROND photometric observations of $z>4$ GRBs concluded that theoretically they are able to detect $A_V>0.5$\,mag bursts but on average they find less dusty bursts. However, with their smaller photometric sample they were not able to completely exclude statistical effects.


It is worth mentioning that sub-mm observations of $z>5$ quasars have revealed large dust masses \citep{beelen06,wang13,willott15,venemanns17}. These high redshift massive quasars are rare objects and not representative of the star-forming galaxy population. And while a few more normal galaxies have been detected in the epoch of reionisation with significant dust \citep{watson15,laporte17}, GRBs probe dust in star-forming regions of galaxies well down the faint end slope of the galaxy luminosity function \citep{tanvir17}.

\subsection{Dust and host galaxy stellar masses}
Several studies have shown that a relation exists between stellar mass and dust attenuation of star-forming galaxies \citep[e.g.,][]{pannella15,alvarez-marquez16,mclure18}. In particular, \citet[][]{dunlop17} found the ratio of obscured to un-obscured star-formation activity to be correlated with the stellar mass for galaxies at $z\sim 2$ (see their Fig. 13), suggesting that the drop-off in dust-obscured star-formation density at high redshifts is due to lower number of high-mass galaxies at those redshifts \citep[see][]{bouwens12,bouwens14,bouwens16}.
  
\begin{figure}
  \centering
{\includegraphics[width=\columnwidth,clip=]{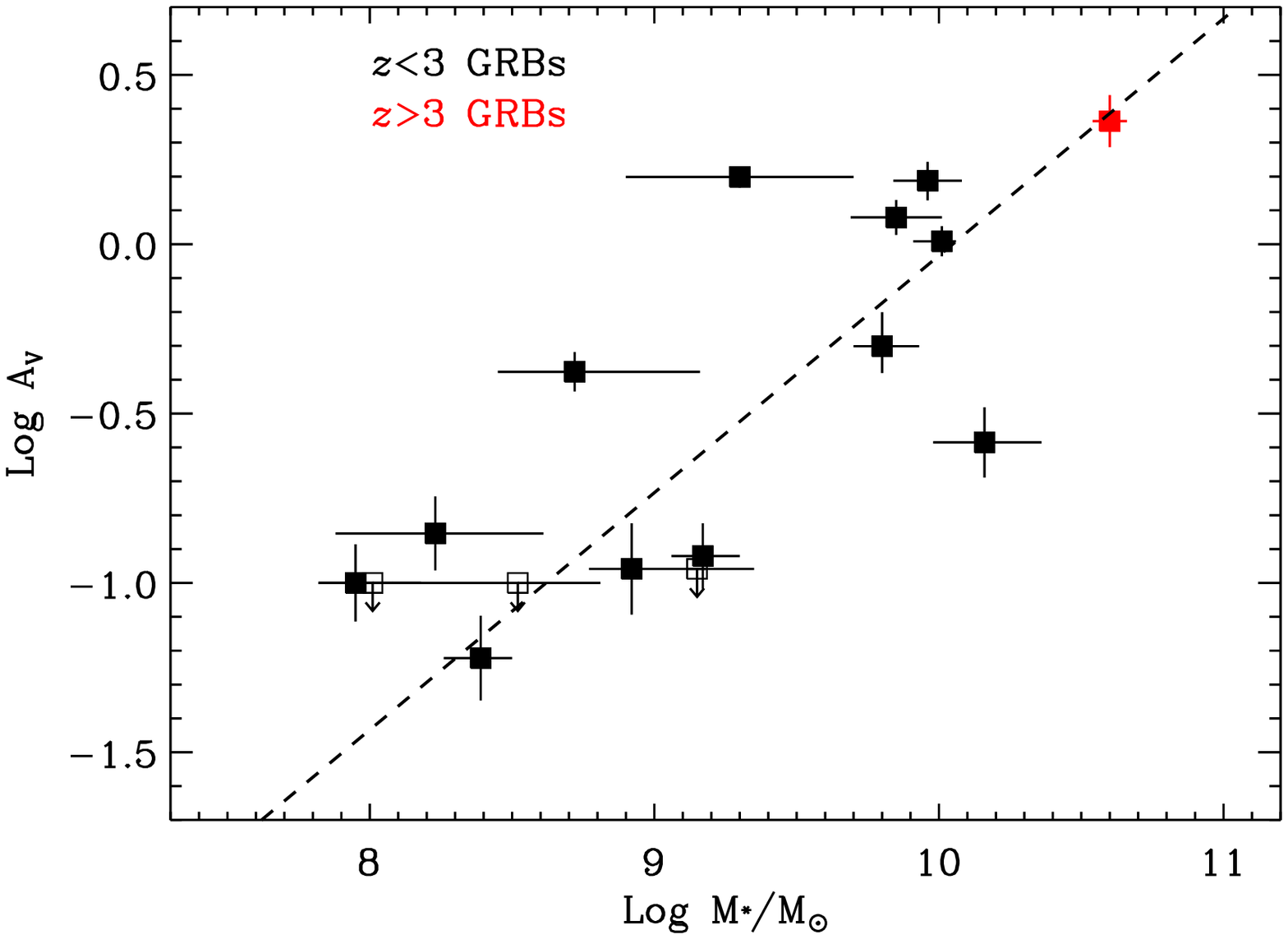}}
     \caption{Visual extinction (single line of sight) against stellar mass, M$_\ast$ for GRBs in our sample. The symbols have the same meaning as in Fig. \ref{nhxav}. The dashed curve corresponds to the  the linear regression fit with Pearson correlation coefficient $r=0.77$ and slope $\alpha=0.70\pm0.05$.}
         \label{smav}
           \end{figure}

We investigate a possible link between the drop in the $A_V$ values of our GRB host sample at $z \gtrsim 3.5$ as reported in \S\ref{dustz} and the redshift dependent galaxy luminosity function which predicts a small number of massive galaxies at those redshifts, we searched the literature for stellar mass measurements of the GRB host sample presented in Fig. \ref{zhist}. These measurements are available for 16 GRB hosts \citep[see \url{http://www.grbhosts.org};][and references therein]{arabsalmani18}, spanning a range between $10^{7.95}<M_\ast/M_\odot<10^{10.60}$ over a redshift range of $0.13<z<3.04$. We find a clear correlation (Pearson correlation coefficient $r=0.77$), with $A_V$ (obtained for single sightline using GRB afterglows) rising with stellar mass from $\sim 0.1$ at $M_\ast/M_\odot = 10^{8.5}$, to $\sim 1.0$ at $M_\ast/M_\odot = 10^{10.5}$ (Fig. \ref{smav}). We therefore conclude that the drop-off in $A_V$ for our GRB hosts at $z \gtrsim 3.5$ could simple reflect that the stellar masses of GRB host galaxies at those redshifts are smaller than those of GRB host galaxies at lower redshifts. \citep[see also][]{tanvir12,mcguire16,corre18}. However, note that GRB host observations are biased against dust.

There is a population of GRBs ($\sim40-50$\%) which do not have identified optical afterglows and are referred to as ``dark'' GRBs \citep[e.g.][]{fynbo09}. This optical bias removes GRBs that are very faint, at very high redshift, or in dusty environments. These `dark' GRBs are found to reside mostly in dusty and massive galaxies galaxies \citep{rossi12,kruhler12b,perley13,kruhler15,perley16}. This also suggests that the dust drop at higher redshift is indicative of the presence of lower stellar mass galaxies at those redshifts.

\subsection{Dust producers}
In the young Universe at $z>4$, dust production in CCSN ejecta is one viable source of dust \citep{todini01,morgan03,marchenko06,hirashita05,dwek07}. However, dust destruction by the SN reverse shock \citep{nozawa07,bianchi07,schneider12,lakicevic15} and a contribution from AGB stars at high redshift \citep{valiante09,hirashita14} is still under debate. Observations of local CCSN remnants suggest dust production of several tenths of a solar mass per remnant, close to maximal dust production from all of the refractory elements synthesised in the core collapse \citep{lucy89,wooden93,sakon09,gomez12,delooze17}. SNe\,Ia, on the other hand, appear to produce no significant dust directly \citep{gomez09}. Finally, there is significant evidence of dust growth in the ISM \citep{jenkins09,decia16}. Whatever the origin of the dust, the elements must first be synthesised.

Evolved, lower mass stars, either as AGB stars or Type\,Ia SNe, are thought to be the major contributors to the two elements that make up at least half the dust mass in the local universe: carbon and iron \citep{gehrz89,dwek07}. The low and intermediate mass AGB stars (with initial masses $0.8\le M/M_\odot \le8$) require $\ge$1\,Gyr to evolve and produce carbon \citep{dwek07}. Similarly, Type\,Ia SNe begin to be important at $z=2-4$ \citep{strolger05}. The first stars -- the so-called Pop\,III stars, free of heavy elements -- are thought to form $\sim200$\,Myr after the Big Bang \citep{bromm99,abel02,karlsson13}.

A debate about the primordial initial mass function means that we still do not know the distribution of stellar masses in the early Universe \citep{gall11}. There could be both low and high ($>$7--8\,M$_\odot$) mass AGB stars present. AGB stars with masses exceeding about 2\,M$_\odot$ are expected to experience hot bottom burning (hereafter HBB, \citealt{siess02,constantino14}) and hence most of the carbon produced during the AGB phase is burned into (primary) nitrogen. However, as the envelope mass decreases, HBB stops but dredge-up continues \citep{frost98, tashibu17} with the result that these massive stars spend a brief period as C-stars near the ends of their lives. Hence even these massive AGB stars can contribute to the carbon content in the early universe. Further, super-AGB stars of very low metallicity can also produce substantial amounts of silicon isotopes (Gil Pons, private communication). Their short lives mean that they can form from the ejecta of early Pop\,III SNe and possibly contribute to the production of silicate dust in the early universe.

Calculations show that primordial stars of 1.5 M$_\odot$ have lifetimes of about 1.6\,Gyr \citep{marigo01,siess02}. These stars become carbon stars during their AGB phase and
 when they reach the end of their lives they produce carbonaceous dust. But more
 importantly they are expected to produce significant amounts of elemental carbon, which
 is required even for non-stellar sources of dust production. Further, at $z<4$ the first Type\,Ia supernovae start to appear. These are expected to be a significant source of iron, a major component of the dust in the ISM.

Thus we hypothesise that the large injection of carbon and iron could be the cause
 of the increase in dust content at lower redshifts. It is at this redshift that the ISM begins to be enriched in ejecta from the first Type\,Ia supernovae, as well as the first stars that do not experience HBB, i.e. the sudden increase in dust content is enabled by new stellar sources of carbon and iron. Specifically we suggest that the carbon is produced by normal, albeit Pop\,III, AGB stars of masses about 1.6--2 M$_\odot$, and the iron is provided by the first Type\,Ia supernovae. One caveat to this hypothesis is that GRB host galaxies are star-forming--dominated galaxies with young stellar ages \citep{schulze15}. However, this does not preclude earlier generations of stars in these galaxies.

It may also be feasible that the transition we observe to greater dust in star-forming regions at $z\lesssim3.5$ and the one detected in dust-emitting galaxies \citep{dunlop17}, is a more gradual process, simply due to increasing overall metallicity in the star-forming galaxies caused by greater numbers of CCSNe.


\section{Conclusions}
In this work, we derive individual extinction curves of GRB afterglows to study dust properties at $z\ge3$. We use a sample of $z\ge3$ GRBs observed with the VLT/X-shooter, finding 10 new cases where simultaneous photometric observations are available. After correcting sub-optimal flux calibration through photometry and generating SEDs, we find that six out of 10 GRBs are dusty. We combine the individual extinction curves of all $z\ge3$ GRBs observed with X-shooter. The mean $z\ge3$ GRB extinction curve is consistent with the SMC-Bar curve from \citet{gordon03}. We compare visual extinctions of spectroscopically-selected GRBs at all redshifts, indicating a decrease at $z\sim3.5$, with no moderately extinguished event. We further check for observational biases using template spectra, up to $z\sim8$ a burst is detectable with an hour of X-shooter time with an $A_V\sim0.3$\,mag with dust content increasing towards lower redshifts. This suggests that the lack of high redshift moderately extinguished GRBs is not due to instrument sensitivity, although there are other observational biases noted. The uniformly low dust values indicate a decrease in dust content for $z > 3.5$ suggesting a transition in the nature of dust producers. Evolved low and intermediate mass AGB stars require $>1$\,Gyr to produce the carbon that plays such an important role in the formation of dust. We postulate that the  dramatic increase in dust content at $z = 3.5$ is enabled by the production of elements from two lower mass stellar sources occurring at the same time and for the first time: carbon from the death of the first AGB stars that are not massive enough for HBB, i.e. from the death of primordial carbon stars, and iron from the first Type\,Ia SNe. Alternatively, the dust content drop at $z \gtrsim 3.5$ could be the result of the low stellar mass of the GRB host galaxies at such redshifts.

\section*{Acknowledgements}
We are thankful to James Dunlop for a useful referee report. TZ is thankful to Maryam Arabsalmani for helpful discussions. The X-ray data for this work are obtained from the UK \emph{Swift} Science Data Center at the University of Leicester. We are thankful to Pilar Gil Pons and Carolyn Doherty for useful discussions. DW is supported by Independent Research Fund Denmark grant DFF - 7014-00017. JJ acknowledges support from NOVA and NWO-FAPESP grant for advanced instrumentation in astronomy.


\bibliographystyle{aa}
\bibliography{xsh-z.bib}{}

\bsp

\label{lastpage}
\end{document}